\documentclass[fleqn,usenatbib]{mnras}

\usepackage[T1]{fontenc}
\usepackage{ae,aecompl}
\usepackage{array}
\usepackage{graphicx}	
\usepackage[nice]{nicefrac}
\usepackage{bm}
\usepackage{textcomp}
\usepackage{hyperref}
\usepackage{natbib}
\usepackage{amsmath}
\usepackage{amssymb}
\usepackage{graphicx}
\usepackage{xcolor}
\usepackage{multirow}
\usepackage{float}
\usepackage{booktabs}
\usepackage{subfig}
\usepackage{lineno}
\usepackage{seqsplit}
\usepackage{verbatim}
\usepackage{ulem}
\usepackage{tikz}
\usepackage{calrsfs}
\DeclareMathAlphabet{\pazocal}{OMS}{zplm}{m}{n}
\usetikzlibrary{positioning}
\usetikzlibrary{shapes,snakes}
\usepackage{amsfonts}
\usepackage{tabularx}
\usepackage{empheq}
\usepackage[many,most,skins]{tcolorbox}
\usepackage{algorithm}
\usepackage{algpseudocode}

\usepackage{color}
\definecolor{green}{rgb}{0.0,0.5,0.0}
\definecolor{darkblue}{rgb}{0.,0.,0.4}
\definecolor{darkred}{rgb}{0.5,0.,0.}
\hypersetup{urlcolor=darkred, citecolor=blue, linkcolor=blue}

\usepackage{mathptmx}

\definecolor{ForestGreen}{RGB}{34, 139, 34}

\newcommand{\bb}[1]{\mathbb{#1}}
\newcommand{\bs}[1]{\boldsymbol{#1}}
\newcommand{\tm}[1]{\textrm{#1}}
\newcommand{\sans}[1]{\textsf{#1}}

\newcommand{\mc}[1]{\pazocal{#1}}

\tcbuselibrary{theorems}

\newtcbtheorem[number within=section]{ournote}{~~~}%
{colback=green!3,colframe=green!55!black,fonttitle=\bfseries, separator sign none}{nt}

\title[Assessment of Gradient-Based Samplers]{Assessment of Gradient-Based Samplers in Standard Cosmological Likelihoods}

\author[A. Mootoovaloo et al]{
Arrykrishna Mootoovaloo $^{1,2}$\thanks{E-mail: arrykrishna.mootoovaloo@physics.ox.ac.uk},
Jaime Ruiz-Zapatero $^{3,4}$,
Carlos Garc\'ia-Garc\'ia $^{2}$,
David Alonso $^{2}$
\\
$^{1}$ LSST-DA Catalyst Fellow\\
$^{2}$ Oxford Astrophysics, Department of Physics, University of Oxford, Denys Wilkinson Building, Keble Road, Oxford, OX1 3RH, UK\\
$^3$ Department of Physics and Astronomy, University College London, Gower Street, London WC1E 6BT, UK. \\
$^4$ Advanced Research Computing Centre, University College London, 90 High Holborn, London WC1V 6LJ, UK.
}

\date{Accepted XXX. Received YYY; in original form ZZZ}

\pubyear{2023}

\begin{document}
\label{firstpage}
\pagerange{\pageref{firstpage}--\pageref{lastpage}}
\maketitle

\begin{abstract}

\noindent 
We assess the usefulness of gradient-based samplers, such as the No-U-Turn Sampler ($\tt{NUTS}$), by comparison with traditional Metropolis-Hastings algorithms, in tomographic $3\times 2$ point analyses. Specifically, we use the DES Year 1 data and a simulated future LSST-like survey as representative examples of these studies, containing a significant number of nuisance parameters (20 and 32, respectively) that affect the performance of rejection-based samplers. To do so, we implement a differentiable forward model using {\tt JAX-COSMO} \citep{2023OJAp....6E..15C}, and we use it to derive parameter constraints from both datasets using the {\tt NUTS} algorithm as implemented in \S\ref{sec:des-year-1}, and the Metropolis-Hastings algorithm as implemented in {\tt Cobaya} \citep{2013PhRvD..87j3529L}. When quantified in terms of the number of effective number of samples taken per likelihood evaluation, we find a relative efficiency gain of $\mc{O}(10)$ in favour of {\tt NUTS}. However, this efficiency is reduced to a factor $\sim 2$ when quantified in terms of computational time, since we find the cost of the gradient computation (needed by {\tt NUTS}) relative to the likelihood to be $\sim 4.5$ times larger for both experiments. We validate these results making use of analytical multi-variate distributions (a multivariate Gaussian and a Rosenbrock distribution) with increasing dimensionality. Based on these results, we conclude that gradient-based samplers such as $\tt{NUTS}$ can be leveraged to sample high dimensional parameter spaces in Cosmology, although the efficiency improvement is relatively mild for moderate ($\mc{O}(50)$) dimension numbers, typical of tomographic large-scale structure analyses.
\end{abstract}


\section{Introduction}
\label{sec:introduction}

Cosmology has witnessed a transformative integration of machine learning methodologies into its toolbox. This has been partially prompted by the growing complexity of cosmological datases coupled with the increasingly intricate nature of the theoretical models needed to describe them. In this sense, emulating techniques have become important in modelling complicated functions in cosmology.

In very simple terms, emulation entails finding an approximate function which can model the quantity we are interested in. The idea of emulation is in fact an old concept. For instance, \citet{1998ApJ...496..605E} derived an analytic expression to describe the linear matter transfer function in the presence of cold dark matter, radiation, and baryons. However, deriving these types of expressions in general, purely in terms of cosmological parameters, requires significant human ingenuity, particularly in the presence of growing model complexity and ever more stringent accuracy requirements. Various types of emulators have been designed, with their own advantages and disadvantages. Techniques such as polynomial regression, neural networks, Gaussian Processes (GPs) and genetic algorithms have been explored by different groups. For example, \citet{2007ApJ...654....2F} used polynomial regression to emulate the CMB power spectra, while \citet{2007PhRvD..76h3503H} used Gaussian Processes, together with a compression scheme, to emulate the non-linear matter spectrum from simulations. Recently, \citet{2021arXiv210414568A,2022MNRAS.511.1771S,2024OJAp....7E..10B} developed a neural network framework to emulate different power spectra. Moreover, \citet{2023arXiv231115865B} used Symbolic Regression -- a technique for finding mathematical expression of the function of interest -- to emulate the matter power spectrum. Finally, \citet{2022A&C....3800508M} introduced a Gaussian-Process-based approach to emulate both the linear and non-linear matter power spectra (and \citet{2020MNRAS.497.2213M} explored the combination of emulation and compression in the context of weak lensing). This methodology is further discussed in Section \ref{sec:emulator}.

Besides accelerating cosmological calculations, the availability of emulators also enables us to exploit differentiable parameter inference methods. These methods, which include Hamiltonian Monte-Carlo (HMC) samplers, and variational inference schemes, exploit the knowledge of the likelihood derivatives to dramatically improve the acceptance rate of the sample in high-dimentional spaces, or to obtain an approximate form for the marginal posterior \citep{1987PhLB..195..216D, 2016arXiv160100670B}. Our goal in this paper is to explore the extent to which differentiable methods can be used to accelerate the task of parameter inference in standard (e.g. power-spectrum-based) cosmological analyses. In particular we will compare two sampling methods: $\tt{NUTS}$, a gradient-based HMC sampler, and the state-of-the-art implementation of the Metropolis-Hastings algorithm in $\tt{Cobaya}$ \citep{2011arXiv1111.4246H, 2002PhRvD..66j3511L}. While both methods aim to efficiently explore parameter space, they differ in their approach to incorporating information about the target distribution. $\tt{NUTS}$ leverages the gradient of the log posterior to dynamically adjust its step size and mass matrix during the warmup phase, allowing it to adapt to the local geometry of the posterior distribution. In contrast, $\tt{Cobaya}$ relies on the covariance of the proposal distribution, which requires prior specification and might not capture the full complexity of the target distribution. Understanding the trade-offs between gradient-based and non-gradient-based sampling methods is crucial for selecting the most appropriate approach depending on the characteristics of the problem at hand. Although related studies have been carried out recently in the literature, our aim in this work is to make as fair a comparison as possible between both sampling approaches, keeping all other aspects of the analysis (e.g. the hardware platform used, the level of parallelisation, the usage of emulators) fixed and equal between both methods. While in this work, we compare gradient versus non-gradient based samplers, \citet{2021MNRAS.508.3589K} recently developed $\tt{zeus}$, a non-gradient based sampler and compared it with another non-gradient based sampler, $\tt{EMCEE}$.

Our contributions in this work are as follows, 1) we integrate an emulator for the linear matter power spectrum in $\tt{JAX-COSMO}$ \citep{2023OJAp....6E..15C} and leverage its existing functionalities for computing power spectra for galaxy clustering and cosmic shear, 2) we take advantage of gradient-based samplers such as $\tt{NUTS}$ to sample the posterior of the cosmological and nuisance parameters using DES Year 1 data \citep{2018PhRvD..98d3526A} and a future LSST-like survey data and 3) we perform an in-depth assessment of whether differentiability is helpful in this context. 

In \S\ref{sec:method}, we describe the Gaussian Process framework used here. In \S\ref{sec:metrics}, we elaborate on the different metrics used to assess the performance of the samplers. In \S\ref{sec:des-year-1}, we use the DES Year 1 data to infer the cosmological and nuisance parameters via emulation and gradient-based sampling techniques, comparing the performance of standard and differentiable sampling methods. Moreover, in \S\ref{sec:analytical}, we investigate how the effective sample size for different samplers varies across as a function of model dimensionality making use of analytical distributions. Finally, in \S\ref{sec:lsst}, we look into the performance gain of differentiable samplers using a future LSST-like dataset, before concluding in \S\ref{sec:conclusion}. In Appendix \ref{sec:sampling}, we also briefly review Hamiltonian Monte Carlo sampling techniques and its variant NUTS.

\section{Method}
\label{sec:method}

In this section, we briefly describe the steps towards building the power spectrum emulator used in this work and the methods used to extract the derivatives of our model.

\subsection{Gaussian Process emulation}
\label{sec:gp}

\begin{figure*}
    \centering
    \subfloat[Redshift distributions of the lens galaxies]{{\includegraphics[height=0.20\textheight]{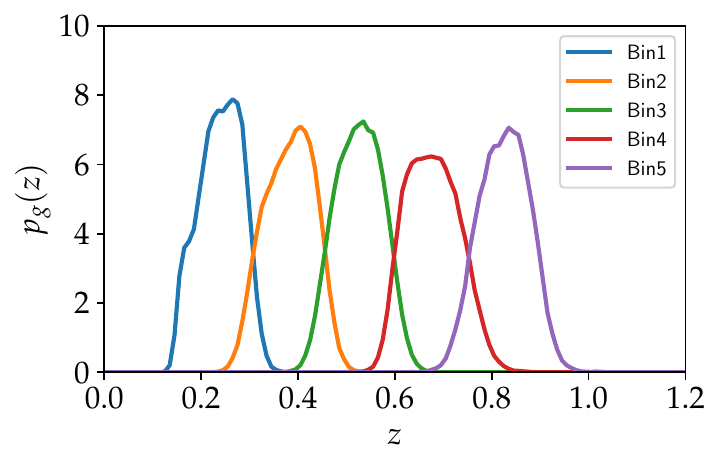}}}
    \qquad
    \subfloat[Redshift distribution of the source galaxies]{{\includegraphics[height=0.20\textheight]{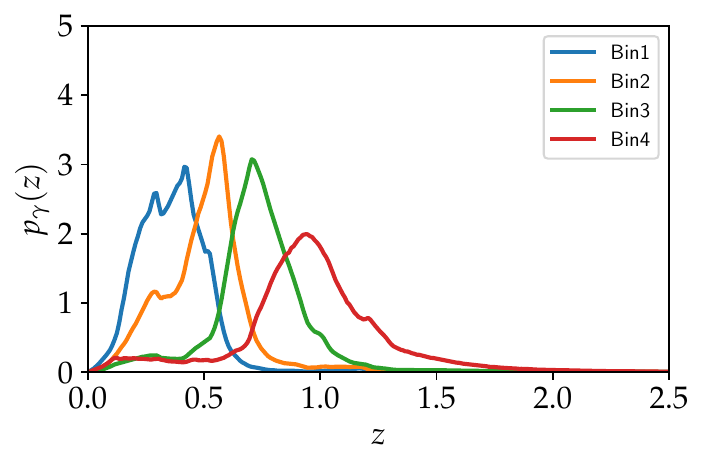}}} 
    \caption{\label{fig:redshift bins}Panel (a) and (b) show the estimated redshift distributions of the lens and source galaxies. In particular, we have five bins for the lens galaxies and four bins for the source galaxies. These bins are used in the calculation of the power spectra for galaxy clustering and cosmic shear. These distributions were estimated and made publicly available by \citet{2018PhRvD..98d3526A}.}
\end{figure*}

A detailed description of Gaussian Processes (GPs) can be found in \citet{2006gpml.book.....R}, and we only outline the methodology briefly here. A GP is essentially a collection of random variables and a finite subset of the variables which has a joint distribution which follows a multivariate normal distribution. It is widely applicable not only to regression problems, but also to active learning scenarios. 

Suppose we have a training set, $\{\sans{X}, \bs{y}\}$. For simplicity, we will assume noise-free regression, such that the data covariance is $\Sigma=\sigma^{2}\bb{I}$, where $\sigma$ is a tiny value, of the order of $10^{-5}$. We will also assume a zero mean prior on the function we want to emulate, $\bs{f}\sim\mc{N}(\bs{0},\,\sans{K})$. $\sans{K}$ is the kernel matrix, for which different functional forms can be assumed. A commonly used kernel is the radial-basis or squared-exponential function (RBF), which is given by:

\begin{equation}
    \sans{K}_{ij}=A\,\tm{exp}\left[-\frac{1}{2}(\bs{x}_{i}-\bs{x}_{j})^{\tm{T}}\Lambda^{-1}(\bs{x}_{i}-\bs{x}_{j})\right]
\end{equation}

\noindent where $A$ is the amplitude of the kernel and $\Lambda$ is typically a diagonal matrix which consists of the characteristic lengthscales for each dimension. Throughout this work, we will use the RBF kernel. 

Given the data and the prior of the function, the posterior distribution of the function can be obtained via Bayes' theorem, that is, 

\begin{equation}    p(\bs{f}|\bs{y},\sans{X})=\dfrac{p(\bs{y}|\sans{X},\,\bs{f})\,p(\bs{f}|\sans{X})}{p(\bs{y}|\sans{X})}
\end{equation}

\noindent where the denominator is the marginal likelihood, also a multivariate normal distribution, $p(\bs{y}|\sans{X})=\mc{N}(\bs{0},\,\sans{K}+\Sigma)$. It is a function of the kernel parameters, $\bs{\nu}=\{A,\,\Lambda\}$. The latter is determined by maximising the marginal likelihood, which is equivalent to minimising the negative log marginal likelihood, that is,

\begin{equation*}
\begin{split}
    \bs{\nu}_{\tm{optimum}} &=\underset{\bs{\nu}}{\tm{arg min}}\hspace{0.15cm} -\tm{log}\,p(\bs{y}|\sans{X})\\
    &=\underset{\bs{\nu}}{\tm{arg min}}\hspace{0.15cm}  \frac{1}{2}\bs{y}^{\tm{T}}(\sans{K}+\Sigma)^{-1}\bs{y}+\frac{1}{2}\tm{log}|\sans{K}+\Sigma| + \tm{constant}.
\end{split}
\end{equation*}

\noindent In general, we are interested in predicting the function at test points in parameter space. For a given test point, $\bs{x}_{*}$, the predictive distribution is another normal distribution with mean and variance given by:
\begin{equation}
    \label{eq:gp_mean_prediction}
    \bar{f}_{*}=\bs{k}_{*}^{\tm{T}}(\sans{K}+\Sigma)^{-1}\bs{y}
\end{equation}

\begin{equation}
    \sigma^{2}_{*}=k_{**}-\bs{k}_{*}^{\tm{T}}(\sans{K}+\Sigma)^{-1}\bs{k}_{*}
\end{equation}

\noindent where $\bs{k}_{*}\in\bb{R}^{N_{\theta}}$ is the kernel computed between the test point and all the training points. Note that the mean is very quick to compute since $\bs{\alpha}=(\sans{K}+\Sigma)^{-1}\bs{y}$ is computed only once and is cached. It then requires $\mc{O}(N_{\theta})$ operations to compute the mean. On the other hand, computing the predictive variance is expensive since it requires $\mc{O}(N_{\theta}^{2})$ operations assuming the Cholesky factor of $\sans{K}+\Sigma$ is cached. 

\begin{table}
\footnotesize
\caption{\label{tab:notations}Symbols and notations with corresponding meanings}
\renewcommand\arraystretch{1.5}
\noindent \begin{centering}
\begin{tabularx}{0.35\textwidth} { 
  | >{\hsize=0.06\textwidth}X 
  | >{\hsize=0.24\textwidth}X | }
\hline
\textbf{Symbol} & \textbf{Meaning}\tabularnewline
\hline 
$N_{\theta}$ & Number of training points\tabularnewline
$N_{k}$ & Number of wavenumbers\tabularnewline
$N_{z}$ & Number of redshifts\tabularnewline
$\bs{x}$ & The data vector of size $N$ \tabularnewline
$\bs{\theta}$ & Inputs to the emulator of size $p$\tabularnewline
$\Theta$ & Input training set of size $N_{\theta}\times p$\tabularnewline
$\bs{y}_{k}$ & Output of size $N_{k}$\tabularnewline
$\bs{y}_{G}$ & Output of size $N_{z}$\tabularnewline
$\sans{Y}_{k}$ & Output training set of size $N_{\theta}\times N_{k}$\tabularnewline
$\sans{Y}_{G}$ & Output training set of size $N_{\theta}\times N_{z}$\tabularnewline
\hline
\end{tabularx}
\par\end{centering}
\end{table}

\begin{figure*}
\noindent \begin{centering}
\includegraphics[width=0.9\textwidth]{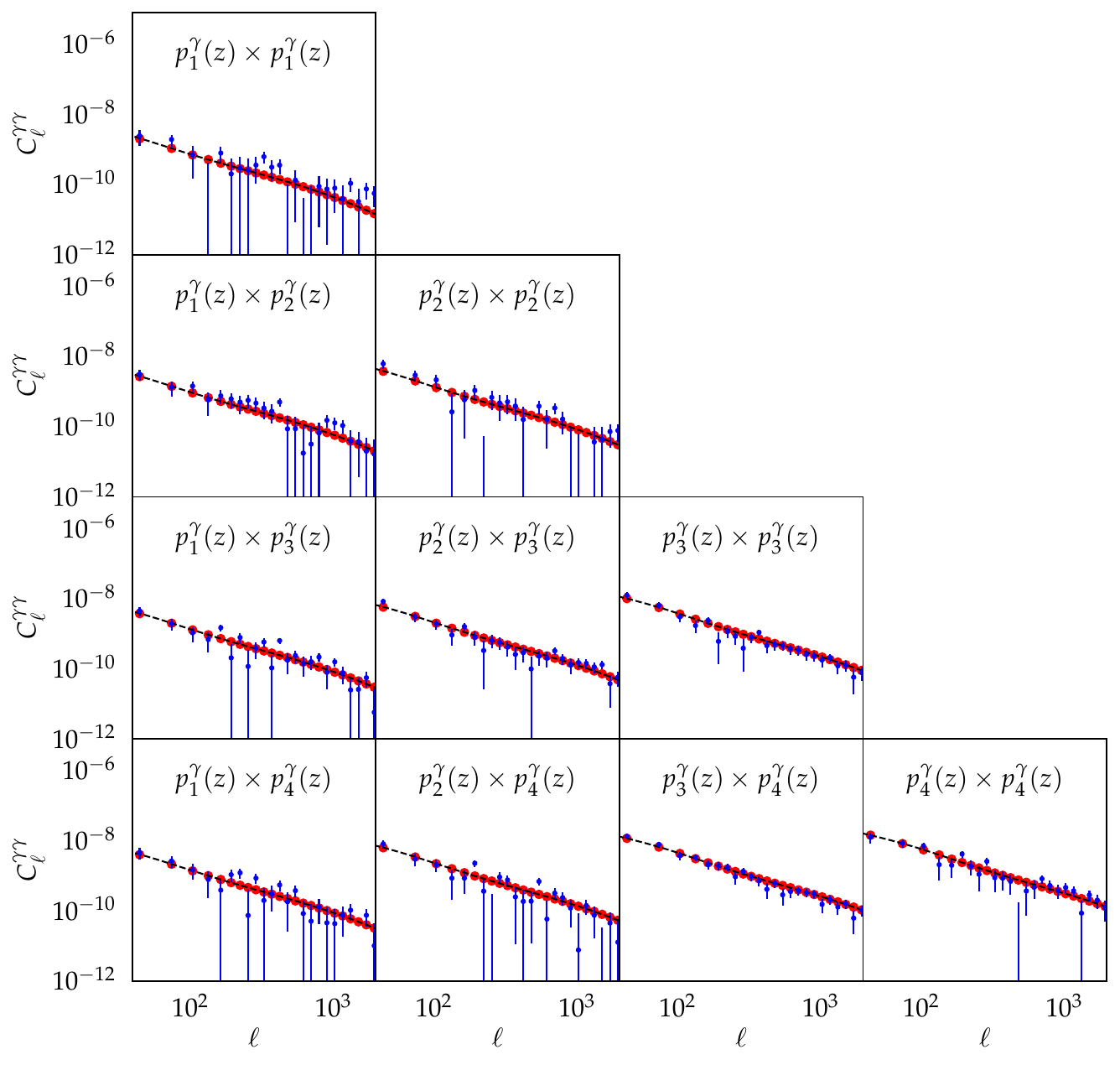}
\par\end{centering}
\caption{\label{fig:shear_data_vector}The data vector for the cosmic shear bandpowers is shown in blue and the red circles show the theorectical bandpowers, where the assumed set of cosmological parameter is $\sigma_{8}=0.841$, $\Omega_{c}=0.229$, $\Omega_{b}=0.043$, $h=0.717$, $n_{s}=0.960$. This corresponds to the mean of the samples obtained when sampling the posterior distribution with $\tt{NUTS}$ and the emulator. The black broken curves show the cosmic shear power spectra. Since we have four tomographic redshift bins (see Figure \ref{fig:redshift bins}), we have a total of 10 auto- and cross-power spectra for cosmic shear.}
\end{figure*}

\subsection{Gradient of the model}
\label{sec:grad_model}
If we define $\bs{\alpha} = (\sans{K}+\Sigma)^{-1}\bs{y}$, the mean function can be written as $\bar{f}_{*}=\bs{k}_{*}^{\tm{T}}\bs{\alpha}$, meaning that, the first derivative of the mean function with respect to the $i^{\tm{th}}$ parameter in the parameter vector $\bs{\theta}_{*}$ (a test point in parameter space) is:

\begin{equation}
    \label{eq:gp_grad_prediction}
    \dfrac{\partial \bar{f}_{*i}}{\partial \bs{\theta}_{*i}} = \left(\dfrac{\partial \bs{k}_{*i}}{\partial \bs{\theta}_{*i}}\right)^{\tm{T}}\bs{\alpha}.
\end{equation}

\noindent As shown in \citet{2022A&C....3800508M}, analytical expression for the first and second derivatives of the mean function can be derived. In this paper, we leverage automatic differentiation functionalities in $\tt{JAX}$ to compute the first derivatives. 

Emulating an intermediate function, for example, the linear matter power spectrum offers various advantages. It has a lower dimensionality, meaning that few training points suffice to accurately model the function. Moreover, once the non-linear matter power spectrum is computed, the calculation of other power spectra such as the shear power spectra involves a rather straightforward linear algebra (integral of the product of the lensing kernel and the matter power spectrum). The latter can be achieved very quickly with existing libraries. 

While the above describes how we can calculate the gradient of the GP with respect to the input parameters, another important quantity is the derivative of the log density of a probability distribution. For example, in a Hamiltonian Monte Carlo sampling scheme, if $f(\bs{\Omega})$ is the target distribution and $\Omega \in \bb{R}^{p}$, it is desirable to have quick access to $-\frac{\partial \tm{ln}f}{\partial \bs{\Omega}}$. A Gaussian likelihood is often adopted in most cosmological data analysis problems. In this spirit, if $\bs{\mu}(\bs{\Omega})$ is the forward theoretical model which we want to compare with the data given a set of parameters $\bs{\Omega}$, and if we define the log-likelihood as $\mc{L}\equiv -\tm{log}\,p(\bs{x}|\bs{\Omega})$, then

\begin{equation}
    \mc{L} = \frac{1}{2}(\bs{x} - \bs{\mu})^{\tm{T}}\sans{C}^{-1} (\bs{x} - \bs{\mu}) + \frac{1}{2}\tm{log}\,|\sans{C}| + \tm{constant}
\end{equation}

\noindent and assuming the covariance is not a function of the parameters, the first derivatives of $\mc{L}$ with respect to $\bs{\Omega}$ is:



\begin{equation}
\label{eq:grad_log_like}
    \dfrac{\partial \mc{L}}{\partial \Omega} = -\dfrac{\partial \bs{\mu}}{\partial \bs{\Omega}}\sans{C}^{-1}(\bs{x} - \bs{\mu}). 
\end{equation}
    
\noindent Note that the derivatives of $\mc{L}$ with respect to $\Omega$ is a $p$ dimensional vector. The above equation implies that we can easily compute the derivatives of the negative log-likelihood if we have access to the derivatives of the forward theoretical model with respect to the input parameters. In general, one can get the derivatives of a complex, expensive and non-linear model by using finite difference method. This is not optimal since for each step in a hybrid Monte Carlo system, one would require $p$ evaluations, assuming a forward-difference or backward-difference method is adopted. Fortunately, with the emulator, approximate derivatives can be obtained (see Equation \ref{eq:gp_grad_prediction}). 

Another quantity of interest is the second derivative of $\mc{L}$ at any point in parameter space. This is essentially the Hessian matrix, 

\begin{equation}
    \sans{H}_{\Omega}\equiv \frac{\partial}{\partial\Omega}\,\left(\frac{\partial \mc{L}}{\partial \Omega}\right)^{\tm{T}}
\end{equation}

\noindent and differentiating Equation \ref{eq:grad_log_like} with respect to $\Omega$, 

\begin{equation}
\label{eq:hessian_log_like}
    \sans{H}_{\Omega}=-\left[\dfrac{\partial}{\partial\Omega}\left(\dfrac{\partial\mu}{\partial\Omega}\right)^{\tm{T}}\right]\sans{C}^{-1}(\bs{x} - \bs{\mu})+\dfrac{\partial\mu}{\partial\Omega}\sans{C}^{-1}\left(\dfrac{\partial\mu}{\partial\Omega}\right)^{\tm{T}}
\end{equation}

\noindent Note that $\sans{H}_{\Omega}\in \bb{R}^{p\times p}$. The parameter vector $\Omega=\hat{\Omega}$ which yields  

$$
\left.\frac{\partial\mc{L}}{\partial\Omega}\right|_{\Omega=\hat{\Omega}}=0,
$$

\noindent corresponds to the maximum likelihood point. For an unbiased estimate of $\Omega=\hat{\Omega}$, the Fisher information matrix is an expectation of the Hessian matrix, that is, $\sans{F}_{\Omega}=\langle\sans{H}_{\Omega}\rangle$. The inverse of the Hessian matrix at this point is the covariance matrix of the parameters. See \citet{1997ApJ...480...22T} for a detailed explanation. 

\section{Metrics}
\label{sec:metrics}
This section describes the different metrics we will use to compare the performance of different samplers, both in terms of their ability to sample the posterior, and of the quality of the resulting parameter chains.

\begin{figure}
\noindent \begin{centering}
\includegraphics[width=0.45\textwidth]{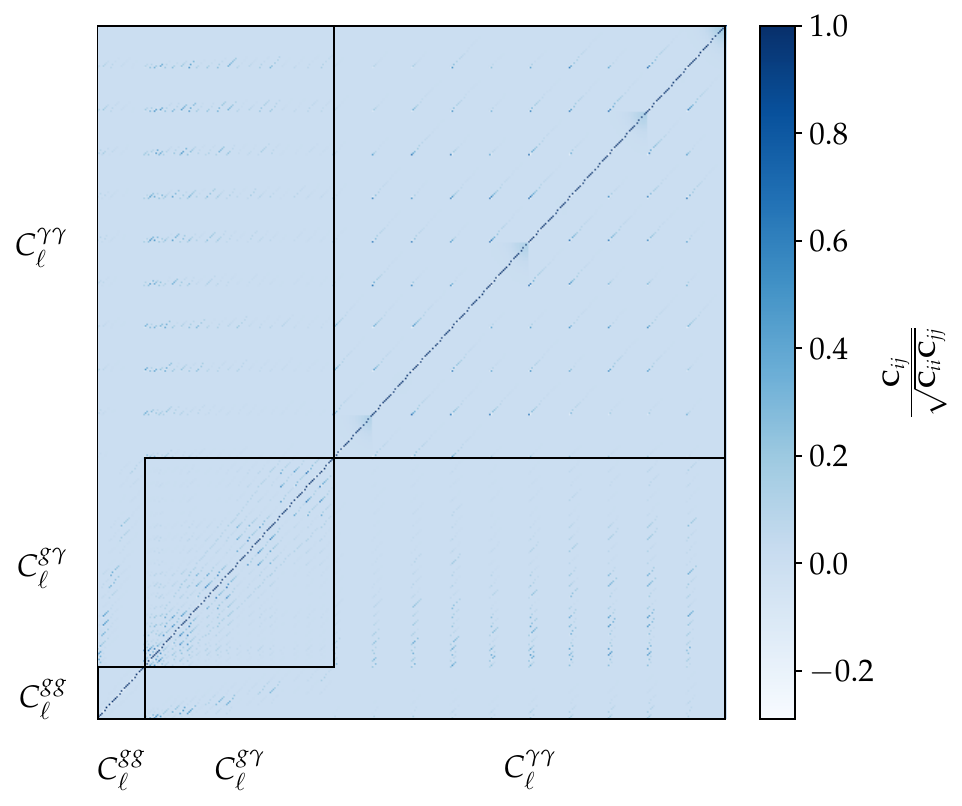}
\par\end{centering}
\caption{\label{fig:data_covariance}The correlation matrix for the galaxy clustering and cosmic shear data. The block matrices along the diagonal show the correlation for the different combinations of probes ($g$ and $\gamma$).}
\end{figure}

\subsection{Kullback-Leibler divergence}
\label{sec:kl}
Suppose we know the exact distribution, $q(\bs{x})\sim\mc{N}(\bs{\mu},\,\sans{C})$ and the distribution recovered by a given sampler $p(\bs{x})\sim \mc{N}(\bs{\bar{x}},\,\hat{\sans{C}})$, and that both of them are multivariate normal. The level of agreement between both distributions can be quantified through the Kullback-Leibler (KL) divergence, which is defined as:
\begin{equation}
\label{eq:dkl}
    D_{\tm{KL}}(p||q)=\int_{-\infty}^{+\infty} p(\bs{x})\,\tm{log}\,\left[\dfrac{p(\bs{x})}{q(\bs{x})}\right]\,\tm{d}\bs{x}.
\end{equation}
For two multivariate normal distributions, this can be computed analytically as:
\begin{equation}
\label{eq:mvn_kl}
    D_{\tm{KL}}=\dfrac{1}{2}\left[\tm{tr}(\sans{C}^{-1}\hat{\sans{C}}) - d + (\bs{\mu} - \bar{\bs{x}})^{\tm{T}}\sans{C}^{-1}(\bs{\mu} - \bar{\bs{x}}) + \tm{ln}\,\dfrac{|\sans{C}|}{|\hat{\sans{C}}|}\right]
\end{equation}
where $d$ is the number of variables. The KL divergence can be understood as a distance measure between distributions. For example, in Equation \ref{eq:mvn_kl}, $D_{\tm{KL}}\rightarrow 0$ as $\bs{\bar{x}} \rightarrow \bs{\mu}$ and $\hat{\sans{C}} \rightarrow \sans{C}$. This metric will only be used for the multivariate normal example in \S\ref{sec:analytical}. Computing the KL divergence, involving non-linear models, will require numerical methods for evaluating high dimensional integrations.

\subsection{Effective Sample Size}
\label{sec:ess}
The Effective Sample Size (ESS) is an approximate measure of the number of effectively uncorrelated samples in a given Markov chain. It is defined as:
\begin{equation}
    n_{\tm{eff}} = \dfrac{m \cdot n}{1+2\sum_{t=1}^{\infty}\rho_{t}},
\end{equation}

\noindent where $m$ is the number of chains, $n$ is the length of each chain, and $\rho_{t}$ is the autocorrelation of a sequence at lag $t$ \citep{gelman_2015}. Suppose we have the samples, $\{\theta_{i}\}_{i=1}^{n}$, the autocorrelation, $\rho_{t}$ at lag $t$ can be calculated using

\begin{equation}
\rho_{t} = \dfrac{1}{\sigma^{2}}\int_{\Theta}\theta^{(i)}\,\theta^{(i+t)}\,p(\theta)\,\tm{d}\theta - \dfrac{\mu^{2}}{\sigma^{2}}
\end{equation}

\noindent for a probability distribution function $p(\theta)$ with mean $\mu$ and variance $\sigma^{2}$. Ultimately, we are interested in the computational cost of each effective sample, since we wish to minimise the time taken to obtain a reasonably sampled posterior distribution. For this reason, in this work we will used a scaled effective sample size, which takes into account the total number of likelihood evaluations:

\begin{equation}
\label{eq:neff}
    N_{\tm{eff}} = \dfrac{n_{\tm{eff}}}{N_{\mc{L}}}
\end{equation}

\noindent where $N_{\mc{L}}$ is the total number of calls of the likelihood function. In a typical (non-gradient) MCMC sampler, this is simply the number of times the posterior probability is evaluated, while in $\tt{HMC}$, this would correspond to the total number of steps performed in all the leapfrog moves.

\subsection{Potential Scale Reduction Factor}
\label{sec:r-stats}
We also report the potential scale reduction factor, $R$ \citep{1992StaSc...7..457G} which measures the ratio of the average of the variance within each chain to the variance of all the samples across the chains. If the chains are in equilibrium, then the value of $\hat{R}$ should be close to one.

Let us assume that we have a set of samples, $\theta_{ij}$, where $i\in[1,n]$ and $j\in[1,m]$. As in the previous section, $n$ is the length of the chain and $m$ is the number of chains. The between-chain variance can be calculated using:

\begin{equation}
    B=\dfrac{n}{m-1}\sum_{j=1}^{m}(\bar{\theta}_{j} - \bar{\theta})^{2}
\end{equation}

\noindent where $\bar{\theta}_{j}=\frac{1}{n}\sum_{i=1}^{n}\theta_{ij}$ and $\bar{\theta}=\frac{1}{m}\sum_{j=1}^{m}\bar{\theta}_{j}$. Furthermore, the within-chain variance is defined as:

\begin{equation}
    W = \dfrac{1}{m}\sum_{j=1}^{m}s_{j}^{2}
\end{equation}

\noindent where $s_{j}^{2}=\frac{1}{n-1}\sum_{i=1}^{n}(\theta_{ij}-\bar{\theta}_{j})^{2}$. The variance estimator of the within- and between- chain variances can be estimated as:

\begin{equation}
    v = \dfrac{n-1}{n}W + \dfrac{1}{n}B
\end{equation}

\noindent Then, the potential scale reduction factor is calculated as:

\begin{equation}
    R = \sqrt{\dfrac{v}{W}}
\end{equation}

\noindent The sampling algorithm is usually started at different initial points in parameter space and in practice, $R-1\sim 0.01$ is deemed to be a reasonable upper limit for convergence to be achieved. 

\section{DES Analysis}
\label{sec:des-year-1}
This section presents our first comparison between HMC and MCMC using, as an example, the 3$\times$2-point analysis of the first-year dataset from the Dark Energy Survey (DES-Y1).

\begin{figure*}
\noindent \begin{centering}
\includegraphics[width=0.90\textwidth]{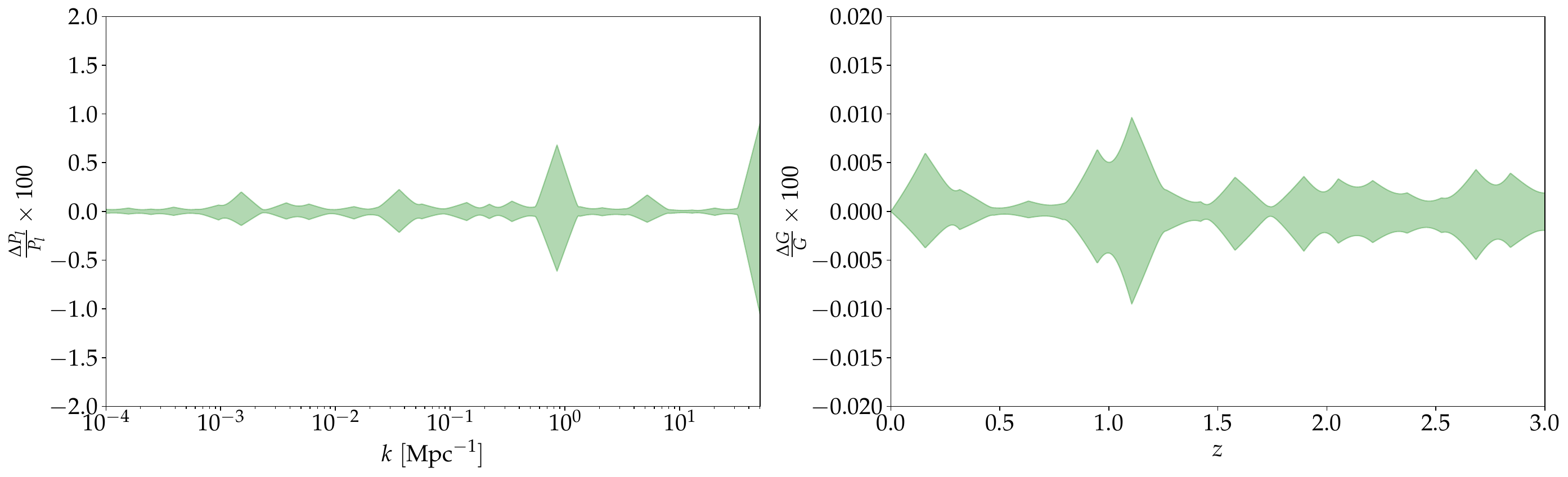}
\par\end{centering}
\caption{\label{fig:emu_accuracy}The left panel shows the accuracy for the linear matter power spectrum, evaluated at $z=0$ over a wavenumber range of $k\in[10^{-4},\,50]$ in units of $\tm{Mpc}^{-1}$ and the right panel shows the accuracy for the quantity, $G$ evaluated over the redshift range of $z\in[0.0\,3.0]$. These quantities can be robustly calculated within an accuracy of $1\%$.}
\end{figure*}

\subsection{Data}
We use the data\footnote{\href{https://github.com/xC-ell/growth-history/tree/main}{https://github.com/xC-ell/growth-history/tree/main}} produced in the reanalysis of DES Y1 carried out by \citet{2021JCAP...10..030G}. The data vector consists of tomographic angular power spectra, including galaxy clustering auto-correlations of the redMaGiC sample in 5 redshift bins, cosmic shear auto- and cross-correlations of the Y1 Gold sample in 4 redshift bins, and all the cross-correlation between clustering and shear samples. The methodology used to construct this data vector was described in detail in \citet{2021JCAP...10..030G}, and will not be repeated here. Following \citet{2021JCAP...10..030G}, we apply a physical scale cut of $k_{\tm{max}}=0.15\,{\rm Mpc}^{-1}$ for galaxy clustering, and angular scale cut $\ell_{\tm{max}}=2000$ for cosmic shear. This leads to a total of 405 data points. 

Figure \ref{fig:shear_data_vector} shows the shear-shear component of the data vector, including the measured bandpowers (blue points with error bars) and the best-fit theoretical prediction (red circles). Moreover, in Figure \ref{fig:data_covariance}, we show the joint correlation matrix for all the data points. The lower block shows the correlation among the data points for galaxy clustering, the middle block along the diagonal for the clustering-shear correlations, and the upper right block for cosmic shear. We assume a Gaussian likelihood of the form

\begin{equation}
    p(\bs{x}|\bs{\theta},\,\bs{\beta}) = \dfrac{1}{\sqrt{|2\pi\sans{C}|}}\,\tm{exp}\left[-\dfrac{1}{2}(\bs{x}-\bs{\mu})^{\tm{T}}\sans{C}^{-1}(\bs{x}-\bs{\mu})\right],
\end{equation}
where $\bs{x}$ is the data vector, $\bs{\mu}(\bs{\theta}, \bs{\beta})$ is the forward theoretical model as a function of the cosmological parameters, $\bs{\theta}$ and the nuisance parameters, $\bs{\beta}$, and $\sans{C}$ is the data covariance matrix.

\subsection{Theory}
\label{sec:theory}

In this section, we discuss the forward model used to model the bandpowers. This also entails a set of nuisance parameters to account for the systematics in the model. Throughout this section, we will denote the normalised redshift distributions for the sources as $p_{\gamma,i}(z)$ and for the lens galaxies as $p_{g,i}(z)$. For example, for the source tomographic distribution $i$,

\begin{equation}
    p_{\gamma,i}(z) = \dfrac{n_{\gamma,i}}{\int n_{\gamma,i}\;dz}
\end{equation}

\noindent where $n_{\gamma,i}(z)$ is the unnormalised distribution.

\subsubsection{Power Spectra}
Assuming a simple bias model for clustering, one which does not depend on redshift, the radial weight function for galaxy clustering in terms of the comoving radial distance, $\chi$ is:

\begin{equation}
    q_{g,i}(\chi) = b_{i}p_{g,i}(z)\,\dfrac{dz}{d\chi}
\end{equation}

\noindent where $b_{i}$ is the galaxy bias. Under the Limber approximation \citep{1953ApJ...117..134L, 2008PhRvD..78l3506L}, the power spectrum for galaxy clustering can be written as:

\begin{equation}
    C_{\ell,i}^{gg}=\int \,d\chi\,\dfrac{q_{g,i}^{2}(\chi)}{\chi^{2}}\,P_{\delta}(k_\ell,z)
\end{equation}

\noindent where $P_{\delta}(k,z)$ is the three-dimensional non-linear matter power spectrum and $k_\ell=(\ell + \nicefrac{1}{2})/\chi$. On the other hand, the lensing efficiency for tomographic bin $i$ is:

\begin{equation}
    q_{\gamma, i} = G_{\ell}\,\dfrac{3H_{0}^{2}\Omega_{m}}{2c^{2}}\,\dfrac{\chi}{a(\chi)}\int_{\chi}^{\chi_{H}}\,d\chi'\,p_{\gamma,i}\,\dfrac{\chi' - \chi}{\chi}
\end{equation}

\noindent where $c$ is the speed of light, $H_{0}$ is the Hubble constant, $a$ is the scale factor and 

\begin{equation*}
    G_{\ell} = \dfrac{\sqrt{\ell(\ell + 2)(\ell + 1)(\ell - 1)}}{(\ell + \nicefrac{1}{2})^{2}}.
\end{equation*}

\noindent The power spectrum due to the correlation of the lens galaxy positions in bin $i$ with the source galaxy shear in bin $j$ is given by:

\begin{equation}
    C_{\ell,ij}^{g\gamma} = (1+m_{j})\,\int \dfrac{q_{g,i}(\chi)\,q_{\gamma,j}(\chi)}{\chi^{2}}\,P_{\delta}(k_\ell,z)\,d\chi
\end{equation}

\noindent where $m_{j}$ is the multiplicative shear bias. Moreover, the shear power spectrum is given by:

\begin{equation}
    C_{\ell,ij}^{\gamma\gamma} = (1+m_{i})(1+m_{j})\,\int \dfrac{q_{\gamma,i}(\chi)\,q_{\gamma, j}(\chi)}{\chi^{2}}\,P_{\delta}(k_\ell,z)\,d\chi
\end{equation}

\noindent Until now, as part of the modelling framework, we have the galaxy bias, $b_{i}$ for galaxy clustering and the multiplicative bias $m_{i}$ for the shear bias. Given that we have five tomographic bins for the lens galaxies and four tomographic bins for the source galaxies (see Figure \ref{fig:redshift bins}), the total number of galaxy biases and multiplicative biases is nine. In the next section, we elaborate on additional nuisance parameters which are accounted for in the modelling framework.

\subsubsection{Intrinsic Alignment and Shifts}
\label{sec:intrinsic_alignment}
The uncertainty in the redshift distributions for both the lens and source galaxies is modelled as a shift in the redshift, that is, 

\begin{equation}
    n_{i}(z)\rightarrow n_{i}(z - \Delta z_{i})
\end{equation}

\begin{table*}
\caption{Priors and inferred values of the cosmological parameters. We report the inferred mean and standard deviation for the different experiments we have run. As anticipated, the inferred mean and standard deviation, irrespective of the two samplers employed, are in close agreement with each other given we are using the same method for computing the different power spectra.\label{tab:cosmo_results}}

\renewcommand{\arraystretch}{1.5}
\noindent \begin{centering}
\begin{tabular}{|l|c|c|c|c||c|c||c|c||c|c|}
\cline{4-11} \cline{5-11} \cline{6-11} \cline{7-11} \cline{8-11} \cline{9-11} \cline{10-11} \cline{11-11} 
\multicolumn{1}{l}{} & \multicolumn{1}{c}{} &  & \multicolumn{4}{c||}{\textbf{NUTS}} & \multicolumn{4}{c|}{\textbf{Cobaya}}\tabularnewline
\hline 
\textbf{Parameters} & $\Theta$ & \textbf{Priors} & $\mu_{\textrm{emu}}$ & $\sigma_{\textrm{emu}}$ & $\mu_{\textrm{EH}}$ & $\sigma_{\textrm{EH}}$ & $\mu_{\textrm{emu}}$ & $\sigma_{\textrm{emu}}$ & $\mu_{\textrm{EH}}$ & $\sigma_{\textrm{EH}}$\tabularnewline
\hline 
Amplitude of density fluctuations & $\sigma_{8}$ & $\mc{U}[0.60,\,1.00]$ & 0.840 & 0.064 & 0.828 & 0.063 & 0.840 & 0.062 & 0.830 & 0.061\tabularnewline
\hline 
CDM density & $\Omega_{c}$ & $\mc{U}[0.07,\,0.50]$ & 0.229 & 0.024 & 0.229 & 0.026 & 0.229 & 0.023 & 0.228 & 0.025\tabularnewline
\hline 
Baryon density & $\Omega_{b}$ & $\mc{U}[0.03,\,0.07]$ & 0.043 & 0.007 & 0.045 & 0.007 & 0.043 & 0.007 & 0.045 & 0.007\tabularnewline
\hline 
Hubble parameter & $h$ & $\mc{U}[0.64,\,0.82]$ & 0.719 & 0.051 & 0.711 & 0.050 & 0.719 & 0.049 & 0.712 & 0.048\tabularnewline
\hline 
Scalar spectral index & $n_{s}$ & $\mc{U}[0.87,\,1.07]$ & 0.957 & 0.056 & 0.958 & 0.056 & 0.958 & 0.054 & 0.959 & 0.054\tabularnewline
\hline 
\end{tabular}
\par\end{centering}
\end{table*}

\noindent hence leading to another set of nine nuisance parameters. The shifts are assumed to follow a Gaussian distribution and we adopt the same priors as employed by \citet{2018PhRvD..98d3526A}. Moreover, we also account for the intrinsic alignment contribution to the shear signal \citep{2011A&A...527A..26J, 2018PhRvD..98d3526A, 2021JCAP...10..030G}. Intrinsic alignments are described through the non-linear alignment model of \cite{2004PhRvD..70f3526H,2007NJPh....9..444B}. In this case, IAs can be simply included as an additive contribution with a radial kernel which is proportional to the tomographic redshift distributions

\begin{equation}
    q_{\tm{IA},i}(\chi) = -G_{\ell}A_{\tm{IA}}(z)p(z)\dfrac{dz}{d\chi}
\end{equation}

\noindent where 

\begin{equation}
    A_{\tm{IA}}(z) = 5\times10^{-14}\,\times\,A_{\tm{IA},0}\,\left(\dfrac{1+z}{1+z_{0}}\right)^{\eta}\,\dfrac{\rho_{c}\Omega_{m}}{D(z)}
\end{equation}

\noindent where $z_{0}=0.62$, $\rho_{c} = 2.775\times 10^{11}\;h^2 \tm{M}_{\odot} \tm{Mpc}^{-3}$ is the critical density of the Universe, $\Omega_{m}$ is the matter density, $D(z)$ is the growth factor. Two nuisance parameters in this model are $A_{\tm{IA}, 0}$ and $\eta$ which are also inferred in the sampling procedure. In short, the total number of parameters in the assumed model is 25, 5 cosmological parameters and 20 nuisance parameters. 

\subsection{Emulation}
\label{sec:emulator}
The default method for computing the linear matter power spectrum in $\tt{JAX-COSMO}$ is the fitting formula derived by \citet{1998ApJ...496..605E}. An alternative approach is to emulate the linear matter power spectrum as calculated by a Boltzmann solver such as CLASS \citep{2011arXiv1104.2932L, 2011JCAP...07..034B}, which we adopt in this work. In particular, using a similar approach developed in \citet{2022A&C....3800508M}, we decompose the linear matter power spectrum in two parts:

\begin{equation}
    P_{l}(k, z, \bs{\theta}) = G(z, \bs{\theta})\,P_{l}^{0}(k,\bs{\theta})
\end{equation}

\noindent where $P_{l}^{0}$ is the linear matter power spectrum evaluated at redshift, $z=0$. The input training points (the cosmological parameters) are generated using Latin Hypercube Sampling (LHS) which ensures the points randomly cover the full space. In particular, the emulator is built over the redshift range of $z\in[0.0,\,3.0]$ and the wavenumber range of $k\in[10^{-4},\,50]$ in units of $\tm{Mpc}^{-1}$. Moreover, we use $N_{\theta} = 1000$ training points according to the prior range shown in Table \ref{tab:cosmo_results}. In particular, we record the targets $(G\tm{ and }P_{l}^{0})$ over $N_{z}=20$ redshift values, equally spaced in linear scale for the redshift range and $N_{k}=30$ wavenumber values, equally spaced in logarithmic scale for the wavenumber range. This gives us two training sets, $\sans{Y}_{k}\in \bb{R}^{N_{\theta} \times N_{k}}$ and $\sans{Y}_{G}\in\bb{R}^{N_{\theta}\times N_{z}}$. We then build 50 independent models as a function of the cosmological parameters. 

As described in \S\ref{sec:gp}, we then use the Gaussian Process formalism to build an emulator which maps the cosmological parameters, $\Theta \in \bb{R}^{N_{\theta}\times p}$, where $N_{\theta}=1000$ and $p=5$ in our application, to each of the target. The models are trained and we store $\bs{\alpha} = (\sans{K}+\Sigma)^{-1}\bs{y}$. During prediction phase, the mean prediction and the first derivative can be calculated using Equations \ref{eq:gp_mean_prediction} and \ref{eq:gp_grad_prediction} respectively. 

One can then use either the emulator or the fitting formula of \citet{1998ApJ...496..605E} to compute the linear matter power spectrum. Moreover, the existing $\tt{JAX-COSMO}$ functionalities are untouched, implying that the non-linear matter power spectrum calculation proceeds via the Halofit fitting formula as a \cite{2012ApJ...761..152T}, which depends directly on the linear matter power spectrum. Furthermore, the shear and galaxy clustering power spectra can be calculated very quickly via numerical integration. 

\subsection{Inference}

In this work, we will compare two samplers namely, $\tt{NUTS}$ and the default sampler in $\tt{Cobaya}$. Throughout this work, we will use the implementation of $\tt{NUTS}$ in $\tt{numpyro}$. In $\tt{numpyro}$, the sampling process is divided into two distinct phases: warmup and sampling.

During the warmup phase, the sampler explores the parameter space to adaptively tune its proposal distribution. It does so by generating a series of samples and adjusting its parameters based on these samples. This phase aims to reach a region of high probability density in the posterior distribution while minimizing the influence of the initial guess. In particular, it focuses on improving the sampler's efficiency by adjusting its step sizes, trajectory lengths, or other parameters to achieve an optimal acceptance rate. The warmup phase is crucial for ensuring that subsequent samples are drawn from the target distribution effectively. In general, if one chooses to adapt the mass matrix during the warmup phase, this can be expensive but once this is learnt, it is fixed and the sampling process can be fast. 

\begin{figure*}
    \centering

    \subfloat[DES Analysis]{{\includegraphics[width = 0.45\textwidth]{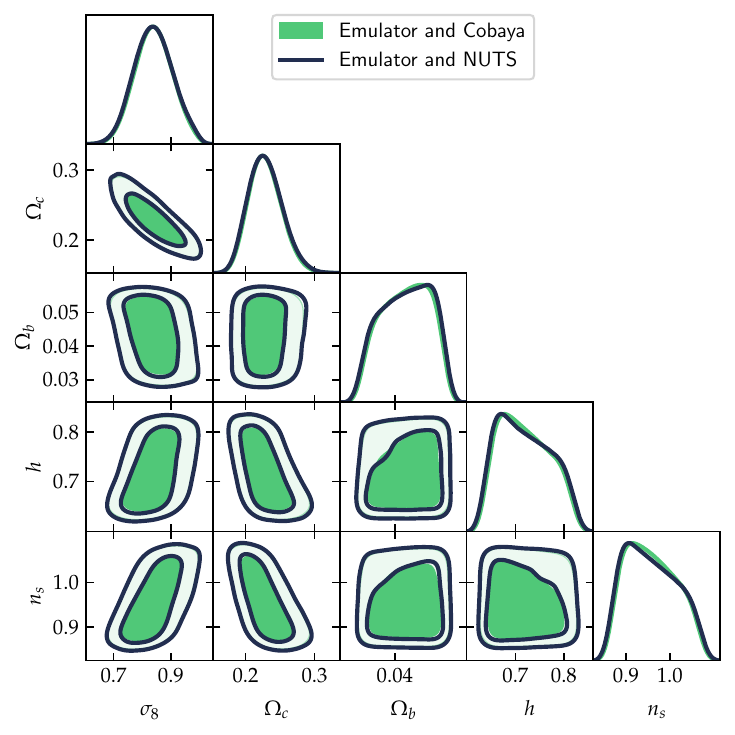}}} 
    \qquad
    \subfloat[Analysis with LSST-like data]{{\includegraphics[width = 0.45\textwidth]{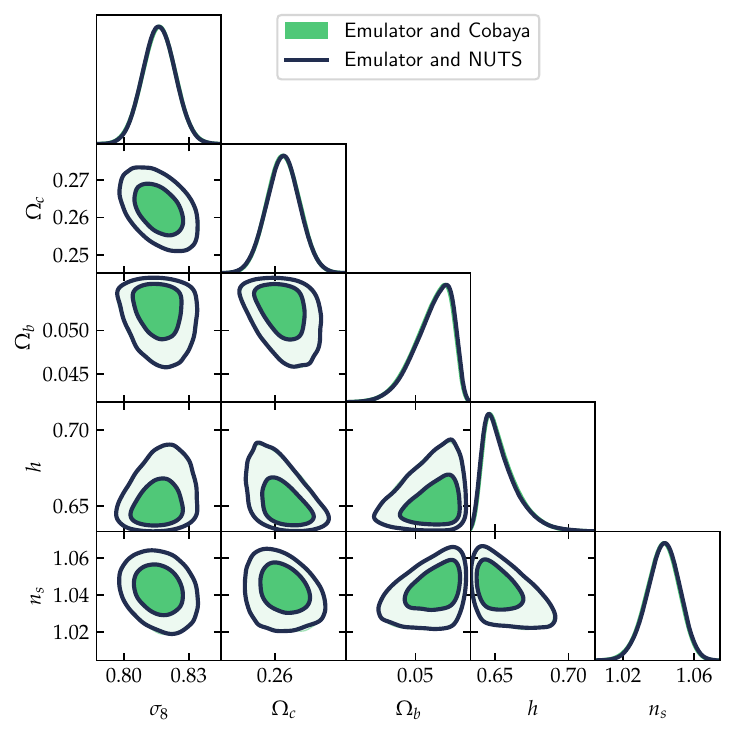}}}
    
    \caption{\label{fig:triangle_plot_cosmo}The 1D and 2D marginalised posterior distribution of all the cosmological parameters. The green contours show the distribution when the emulator is used with the $\tt{Cobaya}$ sampler while the solid black curves correspond to the setup where $\tt{NUTS}$ is used for sampling the posterior distribution. There is negligible difference in the posterior when comparing $\tt{Cobaya}$ and $\tt{NUTS}$. The left panel shows the posterior obtained when using the DES data while the right panel shows the contours obtained with the simulated LSST-like data.}
\end{figure*}

Once the warmup phase is complete, the sampler enters the sampling phase. Here, the sampler generates samples from the posterior distribution according to the adapted proposal distribution. These samples are used for inference and analysis, such as estimating posterior means and variances. 

The warmup phase is essential for the sampler to adapt to the characteristics of the target distribution, while the sampling phase focuses on generating samples for inference. Separating these phases allows $\tt{numpyro}$ to achieve efficient sampling while ensuring the quality of the final samples.

The $\tt{NUTS}$ sampler in $\tt{numpyro}$ involves setting a maximum depth parameter, which determines the maximum depth of the binary trees it evaluates during each iteration. The number of leapfrog steps taken is then constrained to be no more than $2^{j}-1$, where $j$ is the maximum tree depth (see Figure 1 in \citet{2011arXiv1111.4246H}). The sampler reports both the tree depth and the actual number of leapfrog steps computed, along with the parameters sampled during the process. These parameters provide insights into how the sampler explores the parameter space and allow users to monitor the efficiency of the sampling process. Adjusting the maximum depth parameter can help balance exploration and efficiency, ensuring that the sampler explores the posterior distribution effectively while avoiding excessive computational costs. 

For $\tt{NUTS}$, we set the maximum number of tree depth to 8 and use an initial step size of 0.01. Moreover, we fix the number of warmup steps to 500. We also generate two chains consisting of 15000 samples each. 

$\tt{Cobaya}$ uses a Metropolis MCMC as described by \citet{2013PhRvD..87j3529L} and one could also exploit its fast and slow sampler. The idea is to decorrelate the fast and slow parameters so that sampling becomes very efficient. This can result in large performance gains when there are many fast parameters. When using $\tt{Cobaya}$, the sampler stops when either the number of samples specified is attained or the convergence criterion is met $(R-1\leq0.01)$. In some likelihood analyses, it is also possible to adopt an approximate likelihood by analytically marginalising over the many nuisance parameters \citep{2023OJAp....6E..23H, 2023MNRAS.522.5037R}. For $\tt{Cobaya}$, we run two separate chains and we set the number of MCMC samples to $5\times 10^{5}$. 

In all experiments we have performed in this work, we have provided $\tt{Cobaya}$ with a covariance matrix for the proposal distribution. In doing so, we are essentially providing information about how to explore the parameter space efficiently. This vastly helps the sampler to generate samples that are more likely to be accepted, leading to better sampling performance. In principle, one could also specify a mass matrix in $\tt{numpyro}$ for $\tt{NUTS}$. $\tt{NUTS}$ dynamically adapts its step size and mass matrix during the warmup phase. It uses the information from the gradient of the log posterior to tune these parameters. This adaptiveness allows NUTS to efficiently explore the parameter space without requiring explicit specification of the proposal distribution covariance.

\subsection{Results}
\label{sec:des_results}

In this section, we explore the different inference made when using the emulator and the two samplers, $\tt{NUTS}$ and $\tt{Cobaya}$.

In Figure \ref{fig:emu_accuracy}, we show the accuracy of the emulator, which was then embedded into the $\tt{JAX-COSMO}$ pipeline. For the range of redshifts and wavenumber considered and the domain of the cosmological parameters, the quantities $P_{l}$ and $G$ are accurate up to $\leq 1\%$ and $\leq 0.01 \%$ respectively. Recall that we are using 1000 LH samples to build the emulator. Generating the 1000 training points using CLASS took around 1 hour while training the GPs took around 2 hours on a desktop computer. The training of GPs is expensive because of the $\mc{O}(N^{2})$ cost in solving for $\bs{\alpha}=\sans{K}_{y}^{-1}\bs{y}$ via Cholesky decomposition. However, once they are trained and stored, prediction is very fast and computing the log-likelihood is of the order of milliseconds. Moreover, one can use the fixed $\bs{\alpha}$s and the kernel pre-trained hyperparameters in any GP implementation irrespective of whether we use $\tt{numpy}$, $\tt{pyTorch}$, $\tt{TensorFlow}$ and $\tt{JAX}$. Moreover, the priors are sufficiently broad that the emulation framework can be used for different probes, for example, weak lensing as in this context.

Figure \ref{fig:triangle_plot_cosmo} compares the marginalised 1D and 2D distributions of the cosmological parameters, $\bs{\Theta}$ (see Table \ref{tab:cosmo_results}) using the emulator with $\tt{Cobaya}$ and $\tt{NUTS}$. The inferred cosmological parameters are shown in Table \ref{tab:cosmo_results} for different setups.  

Under this configuration, the potential scale reduction factor is equal to 1.00 for all parameters when either the emulator or EH is used in $\tt{JAX-COSMO}$. Sampling the posterior with $\tt{NUTS}$ takes $\sim 13$ hours for two chains with $\tt{numpyro}$ using a single GPU. Alternatively, a single run using $\tt{Cobaya}$, whether with the emulator or EH in $\tt{JAX-COSMO}$, takes approximately 5 hours to sample the posterior. Note that the chains generated by both samplers did not contain the same number of samplers, and therefore the difference in time above is not reflective of their relative performance. Moreover, in order to quantify the difference between the inferred parameters with either sampler, we use the ``difference of Gaussians'' statistic:

\begin{equation}\label{eq:distance_mean_std}
    \delta = \dfrac{|\mu_{\tt{NUTS}} - \mu_{\tt{Cobaya}}|}{\sqrt{\sigma_{\tt{NUTS}}^{2} + \sigma_{\tt{Cobaya}}^{2}}}.
\end{equation}

\noindent The maximum difference among the set of parameters considered in this experiment is $\sim 0.1$. We also compute the average of the scaled effective sample size, $N_{\tm{eff}}$, to compare the samplers. We define the efficiency gain as:

\begin{equation}
\label{eq:gamma_factor}
    \gamma = \dfrac{N_{\tm{eff, \tt{NUTS}}}}{N_{\tm{eff, \tt{Cobaya}}}}.
\end{equation}

\begin{figure}
\noindent \begin{centering}
\includegraphics[width=0.35\textwidth]{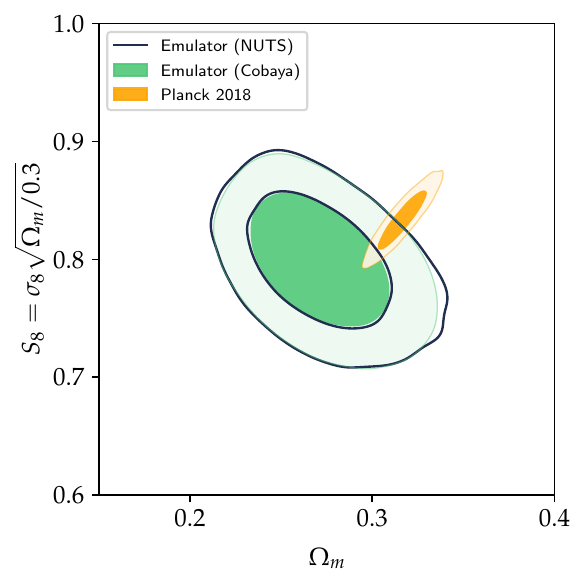}
\par\end{centering}
\caption{\label{fig:s8_tension}The marginalised posterior distribution for the derived parameter, $S_{8}=\sigma_{8}\sqrt{\Omega_{m}/0.3}$ using the emulator. The black contours show the distribution using the sampler in $\tt{NUTS}$ while the green contours correspond to the posterior using $\tt{Cobaya}$. We also plot the Planck 2018 samples (in orange) of the $S_{8}$ and $\Omega_{m}$ which are publicly available.}
\end{figure}

The relative gain in efficiency when using $\tt{NUTS}$ compared to $\tt{Cobaya}$ is $\mc{O}(10)$. When using $\tt{Limberjack}$ and $\tt{NUTS}$ in $\tt{Turing.jl}$, \citet{2023arXiv231008306R} estimated a gain in $N_{\tm{eff}}$ of $\sim 1.7$, compared to the samples obtained using Metropolis-Hastings implemented in $\tt{Cobaya}$ \citep{2021JCAP...10..030G}. In addition, when using reverse mode automatic differentiation (the default setup in $\tt{numpyro}$), the cost of a single gradient calculation to the cost of a single likelihood evaluation is $\sim 4.5$ (either with the emulator or EH). The gain in efficiency is better compared to the cost of the gradient evaluation. With $\tt{julia}$, \citet{2023arXiv231008306R} found this ratio to be $\sim 5.5$ when using forward mode automatic differentiation. Note that the differences in the values above can be attributed to the fact that different samplers will, in general, have different implementations. Taking into account the more expensive gradient evaluation, we find that the overall efficiency gain of $\tt{NUTS}$ with respect to $\tt{Cobaya}$, when measured in terms of computing time on the same platform, is $\sim2$.

In Figure \ref{fig:s8_tension}, we show the joint posterior distribution of the $S_{8}\equiv \sigma_{8}\sqrt{\Omega_{m}/0.3}$ and $\Omega_{m}$ parameters when the posterior is sampled with $\tt{Cobaya}$ and $\tt{NUTS}$. We also use the Planck 2018 samples, which are publicly available, to show the same joint distribution. Note that we are using the baseline $\Lambda$CDM chains with the baseline likelihoods for Planck. With the public Planck chains, $S_{8}=0.834\pm 0.016$. On the other hand, in our experiments, if we use the emulator with $\tt{Cobaya}$ and $\tt{NUTS}$, then $S_{8}=0.797 \pm 0.035$ and $S_{8}=0.797\pm 0.036$. With EH, $S_{8}=0.788\pm 0.033$ and $S_{8}=0.788\pm 0.034$ with $\tt{Cobaya}$ and $\tt{NUTS}$ respectively.

Despite the computational overhead of computing derivatives in $\tt{NUTS}$, we observe a gain in the scaled effective sample size when comparing $\tt{NUTS}$ and $\tt{Cobaya}$. However, it raises the question of whether this advantage will persist in higher-dimensional problems. To address this, we investigate three additional cases: a multivariate normal distribution, the Rosenbrock function, and a future LSST-like system with 37 model parameters. This broader analysis aims to ascertain the scalability and effectiveness of these samplers across various dimensions and problem complexities.

\section{Analytical Functions}
\label{sec:analytical}

In this section, we will investigate how these metrics scale with other functions, such as the multivariate normal distribution and the Rosenbrock function.  

\subsection{Multivariate Normal Distribution}
\label{sec:mvn}
The expression for a multivariate normal distribution is:

\begin{equation}
    q(\bs{x}|\bs{\mu}, \sans{C}) = \dfrac{1}{\sqrt{|2\pi\sans{C}|}}\,\tm{exp}\left[-\dfrac{1}{2}(\bs{x}-\bs{\mu})^{\tm{T}}\sans{C}^{-1}(\bs{x}-\bs{\mu})\right].
\end{equation}

For simplicity, we assume a zero mean and an identity matrix for the covariance of the multivariate normal distribution. The aim is to obtain samples of $\bs{x}$ and to estimate the sample mean, $\bs{\bar{x}}$ and covariance, $\hat{\sans{C}}$ as we increase the dimensionality of the problem. In the limit where we have a large number of unbiased samples of $\bs{x}$, then $\bs{\bar{x}} \rightarrow \bs{\mu}$ and $\hat{\sans{C}} \rightarrow \sans{C}$. 

\subsection{Rosenbrock Function}
\label{sec:rosenbrock}
The next function we consider is the Rosenbrock function, which is given by:

\begin{equation}
    \label{eq:rosenbrock}
    f(\bs{x}) = \sum_{i=1}^{N/2}\left[\zeta(x_{2i-1}^{2}-x_{2i})^{2}+(x_{2i-1}-1)^{2}\right]
\end{equation}

\noindent where $\bs{x}$ is a vector of size $N$ and for this particular variant of the Rosenbrock function, $N$ is even. $\zeta$ is a factor which controls the overall shape of the final function. If it is set to zero, then, the function is simply analogous to the $\chi^{2}$ term in a multivariate normal distribution with mean one and covariance matrix equal to the identity matrix. For $\zeta > 0$, a quartic term is introduced in the overall function and this leads to banana-like posteriors. In both experiments (multivariate normal distribution and the Rosenbrock function), we set the step size and the maximum tree depth to 0.01 and 6 respectively. We use 500 warmup steps and generate two chains consisting of 5000 samples.

We fix $\zeta = 9$ in our experiments. Some of the 2D joint posterior distributions follow a banana-like shapes, which demonstrates the complexity of these functions, especially in high dimensions.

\subsection{Results}
\label{sec:analytical_results}

In Figure \ref{fig:metrics_mvn}, we show how $N_{\tm{eff}}$ changes as we increase the dimensionality of the multivariate normal distribution (in red and blue using $\tt{Cobaya}$ and $\tt{NUTS}$ respectively). When evaluating the effective sample size per likelihood evaluation, Cobaya consistently yields lower values compared to NUTS. This discrepancy arises from their respective sampling strategies. $\tt{NUTS}$ uses the gradient of the likelihood function to build its proposal distribution leading to a higher acceptance rate compared to $\tt{Cobaya}$ whose proposal distribution only depends on the last accepted sample. Consequently, while $\tt{Cobaya}$ requires more likelihood evaluations to achieve a comparable effective sample size to NUTS, the latter tends to provide more informative samples in fewer evaluations. Moreover, Figure \ref{fig:metrics_mvn_distance} shows how the KL-divergence between the inferred distribution and the exact distribution changes as a function of the dimensionality. Interestingly, the KL-divergence is lower for $\tt{Cobaya}$ compared to $\tt{NUTS}$, but the inferred values of the mean and standard deviation are close to zero and unity respectively, with either sampler. The maximum difference between the expected statistics (mean and standard deviation) and the inferred ones across all dimensions is $\sim 0.03$ with either sampler.

\begin{figure}
\noindent \begin{centering}
\includegraphics[width=0.4\textwidth]{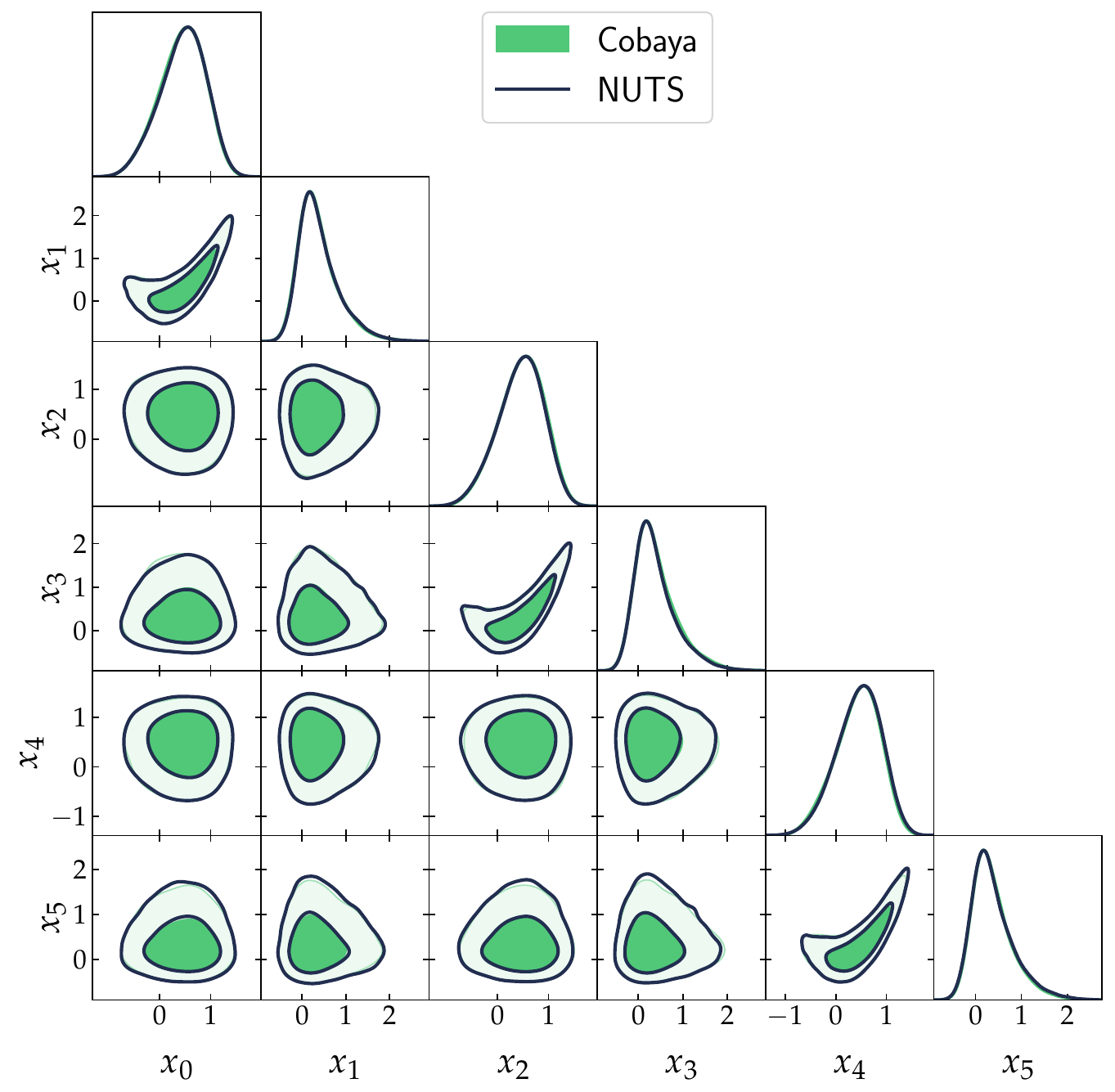}
\par\end{centering}
\caption{\label{fig:rosenbrock_post}The posterior distribution of the first six parameters in a 100 dimensional Rosenbrock function using the $\tt{Cobaya}$ and $\tt{NUTS}$ sampler. In particular, this is obtained by setting $\zeta=9$ in the function (see \S\ref{sec:rosenbrock} for further details).}
\end{figure}

In Figure \ref{fig:metrics_mvn}, we show the scaled effective sample size when we sample the Rosenbrock function with $\tt{NUTS}$ and $\tt{Cobaya}$ (in purple and green respectively) for different dimensions. Although the Rosenbrock function is more complex than the multivariate normal distribution, $N_{\tm{eff}}$ for $\tt{NUTS}$ is superior to that of $\tt{Cobaya}$. Nevertheless, as depicted in the Figure \ref{fig:metrics_mvn}, the $N_{\tm{eff}}$ for the Rosenbrock function is anticipated to be lower than that for the multivariate normal distribution when using the same sampler. In the multivariate normal distribution example, we know the analytic expression for the function and hence, we are able to use the analytic expression for the Kullback-Leibler divergence to estimate the distance between the inferred distribution and the expected distribution. For the Rosenbrock function, this is not the case since we are dealing with a non-trivial function. However, as seen in Figure \ref{fig:rosenbrock_post}, the mean and variance are roughly the same for every two dimensions. In this spirit, we calculate the mean of the difference between the mean of the samples and the expected mean, that is, $\langle \mu_{*} - \mu\rangle$ for all parameters. We do a similar calculation for the standard deviation, that is, $\langle \sigma_{*} - \sigma\rangle$. $\mu_{*}$ and $\sigma_{*}$ are the inferred mean and standard deviation while $\mu$ and $\sigma$ are the expected mean and standard deviation. For $\zeta=9$, $\mu\sim 0.45$ and $\sigma\sim 0.39$ in the first dimension and $\mu\sim 0.39$ and $\sigma\sim 0.48$ in the second dimension. We find that with $\tt{NUTS}$ and $\tt{Cobaya}$, these differences are very close to zero in all dimensions, $\langle \mu_{*} - \mu\rangle \lesssim 0.015$ and $\langle \sigma_{*} - \sigma\rangle \lesssim 0.015$.

Furthermore, in both the multivariate normal distribution and the Rosenbrock function examples, the potential scale reduction factor is close to one when either $\tt{NUTS}$ or $\tt{Cobaya}$ is used to sample the function. However, with a tricky function such as the Rosenbrock function, we find that the potential scale reduction factor gets worse ($R\sim 1.0-1.4$) with $\tt{Cobaya}$ as the dimensionality increases. Moreover, the acceptance probability when $\tt{NUTS}$ is used is always $\gtrsim 0.7$ with either the multivariate normal or the Rosenbrock function. On the other hand, $\tt{Cobaya}$ has an acceptance probability of $\sim 0.3$ when sampling the multivariate normal distribution. With the Rosenbrock function, the acceptance probability varies from $\sim 0.17$ to $\sim 0.1$ as the dimensionality increases. 

Based on the experiments performed in this section, we find that $\tt{NUTS}$ always produces more effective samples, irrespective of the function employed. With the Rosenbrock function depicting non-Gaussianity -- characterized by its non-linear and asymmetric shape -- it is expected that samplers will result in a reduction of $N_{\tm{eff}}$. For $d\leq 100$, both samplers are able to recover the correct shape of the posterior distribution. However, $\tt{NUTS}$ is more likely to scale better to higher dimensions $(d>100)$ as a result of its consistent high $N_{\tm{eff}}$.

\section{LSST}
\label{sec:lsst}
 
Lastly, we look into a $3\times 2$ point analysis using simulated data for a future LSST-like survey. The number of nuisance parameters is expected to be higher in order to account for the astrophysical and observational systematic uncertainties in the shear signal. 

\subsection{Data and Model}
Following the \citet{2018arXiv180901669T} document, we specify ten tomographic spaced by 0.1 in photo-$z$ between $0.2 \leq
z \leq 1.2$  bins for galaxy clustering. For cosmic shear, we assume five tomographic bins \citep{2023OJAp....6E...8L}. We use a physical scale cut of $k_{\rm max}=0.15\,{\rm Mpc}^{-1}$ for galaxy clustering, and a less conservative $\ell_{\rm max}=3000$ for cosmic shear. The simulated data consists of the angular power spectra, $C_{\ell}$, with a Gaussian covariance given by the Knox formula. The $C_\ell$ has been averaged over each of the $\ell$-bins. Both auto- and cross-power spectra are included in the analysis for shear-shear and galaxy-shear correlations and only auto-power spectra are included for galaxy-galaxy correlations. The data vector, after applying the scale cuts, consists of 903 elements, $\bs{x}\in\bb{R}^{903}$ and a data covariance matrix, $\sans{C}\in\bb{R}^{903\times 903}$. In this setup, we now have 

\begin{itemize}
    \item ten bias parameters, $b_{i},\,i\in[1,10]$,
    \item ten shift parameters, $\Delta z_{j}^{g},\,j\in[1,10]$,
    \item five multiplicative bias parameters, $m_{k},\,k\in[1,5]$,
    \item five shift parameters, $\Delta z_{t}^{\gamma},\,t\in[1,5]$ and 
    \item two parameters in the intrinsic alignment model ($A_{IA},\eta)$
\end{itemize}
resulting in a total of 32 nuisance parameters. The total number of parameters in this experiment is 37, the 5 cosmological parameters (shown in Table \ref{tab:cosmo_results}) and 32 nuisance parameters.

\begin{figure}
\noindent \begin{centering}
\includegraphics[width=0.45\textwidth]{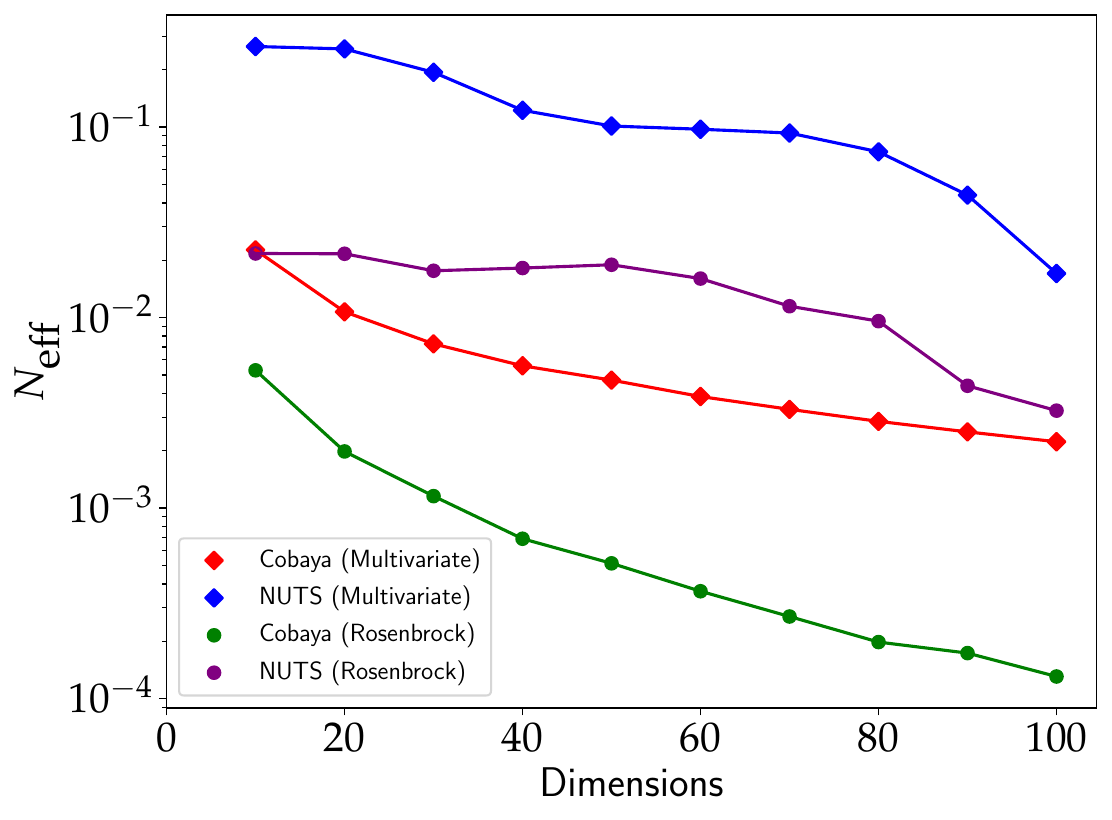}
\par\end{centering}
\caption{\label{fig:metrics_mvn}The figure shows the scaled effective sample size as a function of dimensions of the multivariate normal distribution (red for $\tt{Cobaya}$ and blue for $\tt{NUTS}$) and the Rosenbrock function (green for $\tt{Cobaya}$) and purple for $\tt{NUTS}$). $N_{\tm{eff}}$ for $\tt{NUTS}$ is always higher compared to $\tt{Cobaya}$ over the dimension considered here. Moreover, as expected, for a tricky function such as the Rosenbrock function, $N_{\tm{eff}}$ is always lower compared to the multivariate normal distribution case.}
\end{figure}

In this experiment, for $\tt{NUTS}$, we fix the step size to 0.01 and a maximum tree depth of 8. We also fix the number of warmup steps to 500 and generate two chains of 15000 samples. For $\tt{Cobaya}$, we run two chains and set the number of samples to $5\times 10^{5}$ and the convergence criterion to $R-1\leq0.01$.

\subsection{Results} 
The Gelman-Rubin convergence test, represented by the potential scale reduction factor, consistently indicates convergence to the target distribution for all parameters when employing either $\tt{NUTS}$ or $\tt{Cobaya}$ sampler. This convergence holds true regardless of the forward modelling technique used, whether it be the Eisenstein-Hu method or the emulation method. 

Furthermore, when comparing the $\tt{NUTS}$ and $\tt{Cobaya}$ samplers, the sampling efficiency, $\gamma$ as calculated using Equation \ref{eq:gamma_factor}, is found to be $\mc{O}(10)$. The relative cost of the gradient of the likelihood to the likelihood itself is $\sim 4.5$. Similar to the analysis of the DES Year 1 data (see \S\ref{sec:des-year-1}), the overall gain is $\sim 2$. 

The constraints obtained using $\tt{NUTS}$ and $\tt{Cobaya}$ are comparable, with the maximum distance (see Equation \ref{eq:distance_mean_std}) being $\sim 0.2$ if we use the emulator and $\sim 0.1$ if we use EH method. The marginalised 1D and 2D posterior distribution of the cosmological parameters are shown in right panel of Figure \ref{fig:triangle_plot_cosmo}.

\section{Conclusion}
\label{sec:conclusion}
In this work, we have performed a quantitative assessment of different aspects related to emulation and gradient-based samplers.

In particular, we have integrated an emulator for the linear matter power in $\tt{JAX-COSMO}$. The emulator is both accurate (see Figure \ref{fig:emu_accuracy}) and fast, comparable to the speed-up obtained when using Eisenstein \& Hu method to calculate the linear matter power spectrum. Note that only 1000 training points have been used to achieve an accuracy of $\sim 1\%$, compared to deep learning frameworks, which are generally more data-hungry, and require many training points to achieve almost the same level of accuracy. Moreover, as shown in Figure \ref{fig:triangle_plot_cosmo}, the constraints obtained with different samplers such as $\tt{NUTS}$ and $\tt{Cobaya}$ agree with each other. See Table \ref{tab:cosmo_results} for numerical results. A notable advantage of using emulation technique is that for a particular set of configurations (priors on the cosmological parameters, range of the wavenumber considered and redshift domain), the emulator is trained once, stored and can be coupled to any likelihood code, under the assumption that the scientific problem being investigated does not require configurations beyond those of the emulator. 

In the DES analysis, the use of $\tt{NUTS}$ yields a sampling efficiency gain of approximately 10 for a system with 25 parameters and a non-trivial $\sigma_{8}-\Omega_{c}$ degeneracy. Alongside assessing the effective sample size, the Gelman-Rubin statistics were calculated to confirm convergence across all chains in each experiment. While the efficiency gain favours $\tt{NUTS}$ in this scenario, it is not by a significant margin (roughly 2 when comparing sampling efficiency to the relative cost of gradient calculation). This observation aligns with findings by \citet{2023arXiv231008306R}.   

\begin{figure}
\noindent \begin{centering}
\includegraphics[width=0.44\textwidth]{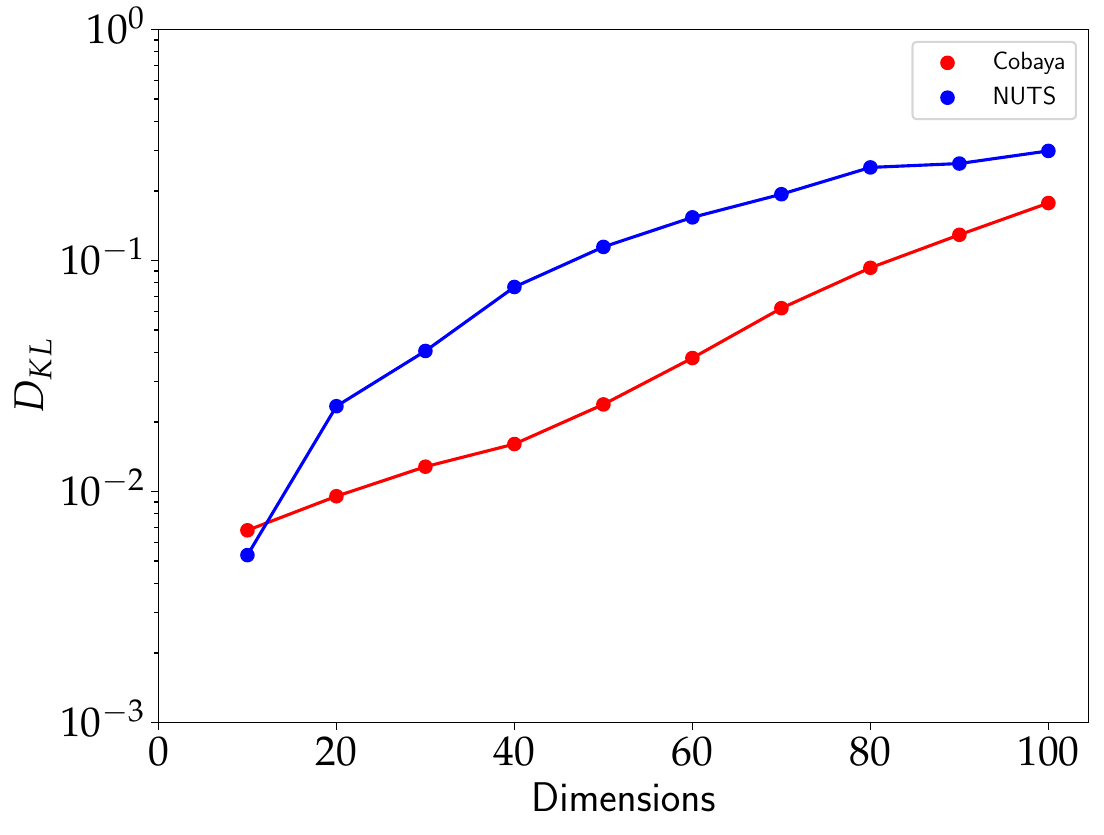}
\par\end{centering}
\caption{\label{fig:metrics_mvn_distance}As elaborated in \S\ref{sec:kl}, we also compute the Kullback-Leibler divergence, $D_{\tm{KL}}$, between the sampled distribution and the expected one. For the dimensions considered here $(10 - 100)$, the $D_{\tm{KL}}$ for $\tt{Cobaya}$ is smaller in this range (except for $d=10$). However, the inferred mean and the standard deviation in either case are close to 0 and 1 respectively.}
\end{figure}

Given the DES analysis, we also investigate the advantages of $\tt{NUTS}$ over standard non-gradient based samplers like $\tt{Cobaya}$ as a function of dimensionality, and we find that $\tt{NUTS}$ is preferable in high dimensions $(d>100)$. The comparison, illustrated with the multivariate normal distribution, reveals several key advantages of $\tt{NUTS}$ in high-dimensional contexts. Firstly, the small Kullback-Leibler divergence indicates that the chain converges to expected results efficiently. Secondly, the Gelman-Rubin statistics remain close to 1.00, indicating convergence across chains. Thirdly, the scaled effective sample size consistently outperforms $\tt{Cobaya}$, demonstrating the effectiveness of $\tt{NUTS}$. This analysis underscores the utility of $\tt{NUTS}$ in efficiently exploring complex parameter spaces, especially in high dimensions where other methods like $\tt{Cobaya}$ may struggle.

Furthermore, an in-depth examination using the Rosenbrock function confirms the sampling efficiency gain of $\tt{NUTS}$. This suggests that $\tt{NUTS}$ is not only advantageous in terms of convergence and effective sample size but also provides improved exploration of complex, non-trivial functions. Overall, these findings highlight the contexts where $\tt{NUTS}$ outperforms traditional non-gradient based samplers, making it a valuable tool for Bayesian inference in a wide range of applications.

In exploring a 37-dimensional parameter inference problem with $\tt{NUTS}$ and $\tt{Cobaya}$ for a future LSST-like survey data, we have also found that $\tt{NUTS}$ is more effective than $\tt{Cobaya}$ by factor of $\sim 2$. $\tt{NUTS}$ consistently exhibits higher sampling efficiency and provides a greater effective sample size per likelihood call compared to $\tt{Cobaya}$. This suggests that, while $\tt{NUTS}$ is better suited for handling complex parameter spaces with $O(40)$ dimensions, the relative improvement factor to be expected on a given platform is mild ($\sim2$ instead of orders of magnitude). While the cost of gradient calculation is a drawback of using $\tt{NUTS}$, this process can be accelerated by leveraging Graphics Processing Units (GPUs). GPUs excel in parallel processing tasks, including gradient computations, offering significant speedup over traditional CPU-based methods. By harnessing GPU acceleration, $\tt{NUTS}$ becomes more feasible for high-dimensional problems and large-scale cosmological analyses. 

In order to fully exploit this possibility, more complete and sophisticated theory prediction frameworks will need to be developed, able to flexibly produce predictions for a wide range of observables of interest to current and future large-scale structure and CMB experiments. Various approaches to this problem have been initiated by the community, making use of tools such as $\tt{JAX}$ \citep{2023OJAp....6E..20P,2023OJAp....6E..15C,2024OJAp....7E..11R,2024arXiv240512965P} and $\tt{Julia}$, and efforts to bring these frameworks to full maturity will significantly improve our ability to obtain both fast and robust parameter constraints from future data.
\newpage
\section*{Acknowledgement}
We thank Dr. Zafiirah Hosenie for reviewing this manuscript and providing useful feedbacks. We thank Prof. Alan Heavens and Prof. Andrew Jaffe for insightful discussions. AM is supported through the LSST-DA Catalyst Fellowship project; this publication was thus made possible through the support of Grant 62192 from the John Templeton Foundation to LSST-DA. JRZ is supported by UK Space Agency grants ST/W002574/1 and ST/X00208X/1. CGG and DA are supported by the Beecroft Trust. We made extensive use of computational resources at the University of Oxford Department of Physics, funded by the John Fell Oxford University Press Research Fund. 

\section*{Data Availability}
The code and part of the data products underlying this article can be found at: \href{https://github.com/Harry45/DESEMU/}{https://github.com/Harry45/DESEMU/}. The full dataset and pre-trained GPs can be made available upon request.

\bibliographystyle{aa}
\bibliography{ads}

\appendix
\section{HMC and NUTS}
\label{sec:sampling}

\subsection{Hamiltonian Monte Carlo}
Hamiltonian Monte Carlo ($\tt{HMC}$) is a sampling technique, developed by \citet{1987PhLB..195..216D} and the method is inspired by molecular dynamics. The motion of molecules follows Newton's law of motions and it can be formulated as Hamiltonian dynamics. In short, we label the parameters governing the full theoretical model the ``position variables'', and introduce an equal number of ``momentum variables''. Starting from an initial state, a final state is proposed, related to the initial state through a Hamiltonian trajectory using the log-posterior as an effective potential, and a kinetic term governed by a mass matrix. A clear benefit of this sampling scheme is that the proposed state has, by construction a high probability of acceptance, due to energy conservation. Moreover, the sampler is now guided by the momentum variables, and is therefore less prone to random transitions. We refer the reader to \citet{2011hmcm.book..113N}, who provides an in-depth and pedagogical review on $\tt{HMC}$.

In cosmology, $\tt{HMC}$ has been adopted in various cases, with the main goal being to sample the posterior distribution of some desired quantities, for example, the power spectrum, in an efficient way. \citet{2007PhRvD..75h3525H} implemented an $\tt{HMC}$ to sample cosmological parameters and found it to be more efficient by a factor of 10, compared to existing MCMC method in \texttt{CosmoMC}. \citet{2008MNRAS.389.1284T} developed an $\tt{HMC}$ sampling scheme to sample the CMB power spectrum while \citet{2019A&A...625A..64J} developed an ambitious sampling framework, Bayesian Origin Reconstruction from Galaxies (BORG) consisting of the $\tt{HMC}$ sampler, to sample over million of parameters. Most recently, \citet{2023OJAp....6E..15C},  \citet{2024OJAp....7E..11R} and \citet{2024arXiv240512965P} have developed auto-differentiable frameworks to analyse angular power spectra with the aim of enabling gradient-based samplers. 

Here, we describe an implementation of the $\tt{HMC}$ algorithm tailored towards using the mean and the first derivative. The second derivative can also be used to tune the mass matrix but one could also resort to using a small MCMC chain to estimate it. In this section, we are assuming that one is updating the mass matrix as we sample the parameters of the system using the second derivative. Suppose $\Omega\in\bb{R}^{d}$ is a $d-$dimensional position vector (the parameters of interest) and $\bs{q}\in\bb{R}^{d}$ is a $d-$dimensional momentum vector. The full state space has $2d$ dimensions and the system is described by the Hamiltonian: 

\begin{equation}
    \mc{H}(\Omega,\bs{q}) = \mc{U}(\Omega) + \mc{K}(\bs{q}).
\end{equation}

\noindent In Bayesian applications, the potential energy is simply the negative log-posterior, that is, 

\begin{equation}
    \mc{U}(\Omega) = -\left[\tm{log}\,p(\bs{x}|\Omega) +\tm{log}\,\pi(\Omega)\right]
\end{equation}

\noindent while the kinetic energy term is 

\begin{equation}
    \mc{K}(\bs{q}) = \frac{1}{2}\,\bs{q}^{\tm{T}}\sans{M}^{-1}\bs{q} + \frac{1}{2}\tm{log}|\sans{M}| + \tm{constant}
\end{equation}

\noindent Note that the determinant of the mass matrix is not ignored in this variant of $\tt{HMC}$, since the mass matrix is updated after every leapfrog move. In other words, the mass matrix is in fact a function of position as we sample the parameter space. The partial derivatives of the Hamiltonian control the evolution of the position and momentum variables, that is, 

\begin{equation}
    \begin{split}
        \dfrac{d\Omega_{i}}{dt}&=\dfrac{\partial \mc{H}}{\partial \bs{q}_{i}}\\
        \dfrac{d\bs{q}_{i}}{dt} &= -\dfrac{\partial \mc{H}}{\partial \Omega_{i}}
    \end{split}
\end{equation}

\begin{algorithm}
\caption{The leapfrog algorithm}\label{alg:leapfrog}
\begin{algorithmic}
\Procedure{leapfrog}{$\Omega_{0}, \bs{q}_{0}$, $N_{L}$, $\epsilon$, $\sans{M}$}
\State $\bs{q} = \bs{q}(t)-\epsilon\frac{\partial \mc{U}}{\partial \Omega}(\Omega_{0})/2$
\For{$i:1\rightarrow N_{L}$}
    \State $\Omega\gets\Omega+\epsilon\sans{M}^{-1}\bs{q}$
    \State $\bs{q}\gets \bs{q}-\epsilon \frac{\partial \mc{U}}{\partial \Omega}(\Omega)$
\EndFor
\State $\bs{q}=\bs{q}-\epsilon\frac{\partial\mc{U}}{\partial \Omega}(\Omega)/2$
\State \textbf{return} $(\Omega, -\bs{q})$
\EndProcedure
\end{algorithmic}
\end{algorithm}

\noindent Crucially, as discussed by \citet{2011hmcm.book..113N}, three key aspects of this formulation are: 1) the Hamiltonian dynamics are reversible, 2) the Hamiltonian is preserved and 3) the volume is preserved. In order to solve the system of differential equations, one typically resorts to numerical techniques, for example, Euler's method. A better alternative is the leapfrog algorithm, which is summarised below:

\begin{equation*}
\label{eq:leapfrog}
\begin{split}
    \bs{q}(t+\frac{\epsilon}{2}) &= \bs{q}(t)-\frac{\epsilon}{2}\frac{\partial \mc{U}}{\partial \Omega}\left[\Omega(t)\right]\\
    \Omega(t+\epsilon) &= \Omega(t) + \epsilon\sans{M}^{-1}\bs{q}(t+\frac{\epsilon}{2})\\
    \bs{q}(t+\epsilon) &= \bs{q}(t+\frac{\epsilon}{2}) - \frac{\epsilon}{2}\frac{\partial \mc{U}}{\partial \Omega}\left[\Omega(t+\epsilon)\right]
\end{split}
\end{equation*}

\noindent where $\epsilon$ is the step size parameter. The leapfrog algorithm is shown in Algorithm \ref{alg:leapfrog}. In particular, one takes a half-step in the momentum at the beginning before doing $N_{L}$ full steps in the momentum and position variables, followed by a final half-step in momentum. The new proposed state is accepted according to Algorithm \ref{alg:acceptance_criteria}. The full $\tt{HMC}$ algorithm is presented in Algorithm \ref{alg:hmc}. 

\begin{algorithm}
\caption{The acceptance criterion}\label{alg:acceptance_criteria}
\begin{algorithmic}
\Procedure{criterion}{$\Omega_{0}$, $\bs{q}_{0}$, $\Omega$, $\bs{q}$}
\State $u\sim \mc{U}[0,1]$
\If{$u<\tm{exp}[\mc{H}(\Omega_{0}, \bs{q}_{0}) - \mc{H}(\Omega, \bs{q})]$}
    \State \textbf{return} $\Omega$
\Else
    \State \textbf{return} $\Omega_{0}$
\EndIf
\EndProcedure
\end{algorithmic}
\end{algorithm}

To ensure a good performance with the $\tt{HMC}$ method, an appropriate mass matrix, step size and number of leapfrog moves are recommended. We briefly touch upon these concepts below. 

\subsubsection{Mass Matrix}

If we can calculate the Hessian matrix, the latter can be used to estimate the mass matrix in the $\tt{HMC}$. The inverse of this matrix, $\sans{H}^{-1}$, gives an estimate of the covariance of the parameters at any point in parameter space. In fact, this covariance is used in the leapfrog algorithm to take a step in the parameters. In \S\ref{sec:step_size_hmc}, we will look into how we can use this Hessian matrix to set the mass matrix, in order to ensure efficient sampling. Note that we do not adapt the mass matrix at every step of the sampling procedure. Otherwise, this will be very expensive. For example, in the $\tt{numpyro}$ library, during the warmup phase, one can adapt the mass matrix and then the learned mass matrix is fixed throughout the sampling process \citep{phan2019composable}. 

\subsubsection{Step size}
\label{sec:step_size_hmc}
In the leapfrog algorithm, we also have to specify the step size, $\epsilon$. If we assume that the posterior distribution of the parameters are roughly Gaussian, then the derivative of the potential energy is approximately $\sans{H}\Tilde{\Omega}$, where $\Tilde{\Omega}$ is the difference between any point in parameter space and the mean. Hence, we can write a single application of the leapfrog method as:
\begin{equation}
\left(\begin{array}{c}
\Tilde{\Omega}(t+\epsilon)\\
\bs{q}(t+\epsilon)
\end{array}\right)=\sans{Q}\,\left(\begin{array}{c}
\Tilde{\Omega}(t)\\
\bs{q}(t)
\end{array}\right)
\end{equation}

\noindent where 

\begin{equation*}
    \sans{Q} = \left(\begin{array}{cc}
\bb{I}-\frac{\epsilon^{2}}{2}\sans{M}^{-1}\sans{H} & \epsilon \sans{M}^{-1}\\
\frac{\epsilon^{3}}{4}\sans{H}\sans{M}^{-1}\sans{H}-\epsilon \sans{H} & \bb{I}-\frac{\epsilon^{2}}{2}\sans{M}^{-1}\sans{H}
\end{array}\right).
\end{equation*}

\noindent The leapfrog move will only be stable if the eigenvalues of the matrix $\sans{Q}$ have squared magnitude of the order of unity. We can write the following characteristic equation:

\begin{equation}
\label{eq:characteristic_equation}
    \left|\lambda^{2}\bb{I}-2\lambda\left(\bb{I}-\frac{\epsilon^{2}}{2}\sans{M}^{-1}\sans{H}\right)+\bb{I}\right| = 0
\end{equation}
\noindent and from the above equation, we can remove the dependence of $\epsilon$ on $\sans{M}$ by setting it to the Hessian matrix, that is, $\sans{M}=\sans{H}$. Note that one would maximally decorrelate the target distribution if the mass matrix is set to the Hessian matrix. This can be seen from the duality and volume preservation of the $(\Omega, \bs{q})$ phase space. A transformation of the kinetic energy (the momentum variables) leads to an equivalent change in the potential energy (the position variables). 

In 1D, if the mass matrix is set to unity, \citet{2011hmcm.book..113N} argues that a step size of $\epsilon < 2\sigma$, where $\sigma$ is the width of the distribution, leads to stable trajectory. In higher dimensions, assuming we have set the mass matrix to the Hessian matrix, solving the characteristic equation (Equation \ref{eq:characteristic_equation}), leads to 

\begin{equation}
    \lambda^{2}-2\left(1-\frac{\epsilon^{2}}{2}\right)\lambda+1=0.
\end{equation}

\noindent When $\epsilon < 2$, this leads to stable trajectories. In practice, we would set $0<\epsilon < 2$. Alternatively, libraries such as $\tt{numpyro}$ have the option to adapt the step size during the warmup phase via a Primal Dual Averaging (PDA) scheme \citep{phan2019composable, bingham2019pyro}. 


\begin{algorithm}
\caption{The Hamiltonian Monte Carlo algorithm}\label{alg:hmc}
\begin{algorithmic}
\Procedure{hmc}{$\Omega_{0}$, $\epsilon$, $N_{L}$, $N$, $\sans{M}$}
\State $i=0$
\State $\bs{q}_{0}\sim\mc{N}(\bs{0},\sans{M})$
\State $S=\{\}$
\While{$i<N$}
    \State $(\Omega,\,\bs{q})\gets\tm{LEAPFROG}(\Omega_{0},\,\bs{q}_{0},\,N_{L},\,\epsilon,\,\sans{M})$
    \State $\Omega\gets\tm{CRITERION}(\Omega_{0},\,\bs{q}_{0},\,\Omega,\,\bs{q})$
    \State $S\gets S\cup \{\Omega\}$
    \State $\bs{q}_{0}\sim\mc{N}(\bs{0},\,\sans{M})$
    \State $\Omega_{0}\gets \Omega$
    \State i++
\EndWhile
\State \textbf{return} $S$
\EndProcedure
\end{algorithmic}
\end{algorithm}

A very small value of $\epsilon$ would increase the number of steps (and therefore of posterior gradient evaluations), while a large value may lead to numerically unstable trajectories. A similar prescription for choosing the mass matrix and setting the step size, is provided in the appendix of \citet{2008MNRAS.389.1284T}, where their main objective was to sample the CMB power spectrum.

\subsubsection{Number of leapfrog moves}

Another parameter which we have to set is the number of leapfrog moves. There is no straightforward way to choose this parameter. We would prefer to avoid periodic trajectories that end close to the starting position. We would also like to minimise the number of gradient computations, in order to reach the pre-specified number of samples earlier. If a bad step size is chosen, the Hamiltonian grows with the number of leapfrog moves and the probability of acceptance decreases considerably. 

\subsection{No-U-Turn Sampler}

The No-U-Turn Sampler ($\tt{NUTS}$) has been designed to circumvent the need to tune the step size, $\epsilon$ and the number of leapfrog moves, $N_{L}$. 

In order to adaptively adjust the step size during the sampling process, $\tt{NUTS}$ uses a technique referred to as Primal Dual Averaging (PDA). PDA helps address this challenge by adaptively tuning the step size based on the trajectory of the sampler. It does so by maintaining a running average of the gradients of the log posterior over the course of the sampling process. This average is used to update the step size such that it scales appropriately with the curvature of the log posterior. The primal-dual averaging algorithm ensures that the step size adjusts smoothly and efficiently, allowing $\tt{NUTS}$ to explore the parameter space effectively without requiring manual tuning from the user. By continuously updating the step size based on the local geometry of the posterior distribution, $\tt{NUTS}$ with PDA can achieve faster convergence and higher sampling efficiency compared to fixed-step-size methods.

As discussed in the previous section, selecting the number of leapfrog moves can be tricky and the challenge is to find a metric to determine if the trajectory is too short, too long or just right. $\tt{NUTS}$ dynamically selects the trajectory length during its sampling process based on the "No-U-Turn" criterion. When $\tt{NUTS}$ starts exploring the parameter space from a given starting point, it extends its trajectory by recursively building a binary tree of states. At each step, $\tt{NUTS}$ evaluates whether to continue extending the trajectory in a particular direction or to stop. This decision is guided by the no-U-turn criterion, which checks whether the sampler is doubling back on itself. The trajectory length is determined dynamically as $\tt{NUTS}$ extends the tree. If the sampler is still exploring promising regions of the posterior distribution, it continues to double the trajectory length. However, if the sampler starts to turn back, indicating that it has likely overshot a relevant part of the distribution, the trajectory is truncated. By dynamically adjusting the trajectory length based on the no-U-turn criterion, $\tt{NUTS}$ ensures that it explores the parameter space efficiently without needing to specify the number of steps in advance. This adaptive behaviour allows $\tt{NUTS}$ to effectively explore complex and high-dimensional distributions, leading to faster convergence and higher sampling efficiency compared to fixed-length trajectory methods. For an in-depth explanation of the $\tt{NUTS}$ algorithm, we refer the reader to \cite{2011arXiv1111.4246H}.

\bsp	
\label{lastpage}
\end{document}